%% file: report.tex
\documentclass[conference,10pt]{IEEEtran}
\usepackage[utf8]{inputenc}
\usepackage[english]{babel}
\usepackage[T1]{fontenc}

\usepackage{url}

\usepackage{tikz}
\usepackage{pgfplots}
\pgfplotsset{compat=newest}
\usepackage{grffile}
\newlength\fheight
\newlength\fwidth

\usepackage{booktabs}
\usepackage{multicol}

\usepackage{caption}
\usepackage[acronym]{glossaries}
\newacronym{lte}{LTE}{Long-Term Evolution}
\newacronym{rach}{RACH}{Random Access Channel}
\newacronym{prach}{PRACH}{Physical Random Access Channel}
\newacronym[plural=eNBs]{enb}{eNB}{Base Station}
\newacronym[plural=UEs]{ue}{UE}{User Equipment}
\newacronym{ml}{ML}{Machine Learning}
\newacronym{lr}{LR}{Logistic Regression}
\newacronym[plural=NNs, firstplural=Neural Networks]{nn}{NN}{Neural Network}
\newacronym[plural=SNRs, firstplural=Signal-to-Noise Ratios]{snr}{SNR}{Signal-to-Noise Ratio}
\newacronym{awgn}{AWGN}{Additive White Gaussian Noise}
\newacronym{zc}{ZC}{Zadoff-Chu}
\newacronym{cp}{CP}{Cyclic Prefix}

\begin{document}

\title{Enabling LTE RACH Collision Multiplicity \\ Detection via Machine Learning}
\author{\IEEEauthorblockN{Davide Magrin\IEEEauthorrefmark{2}, Chiara Pielli\IEEEauthorrefmark{2}, \v{C}edomir Stefanovi\'c\IEEEauthorrefmark{1}, and Michele Zorzi\IEEEauthorrefmark{2}}
\IEEEauthorblockA{\IEEEauthorrefmark{2}Department of Information Engineering, University of Padova, Padova, Italy. Email: {\tt\{name.lastname\}@dei.unipd.it} \\
\IEEEauthorrefmark{1}Department of Electronic Systems, Aalborg University, Aalborg, Denmark. Email: \tt cs@es.aau.dk}}
\maketitle

\begin{abstract}
  The collision resolution mechanism in the \gls{rach} procedure of the
  \gls{lte} standard is known to represent a serious bottleneck in case of machine-type
  traffic. Its main drawbacks are seen in the facts that \glspl{enb} typically
  cannot infer the number of collided \glspl{ue} and that collided \glspl{ue}
  learn about the collision only implicitly, through the lack of the feedback in
  the later stage of the \gls{rach} procedure. The collided \glspl{ue} then
  restart the procedure, thereby increasing the \gls{rach} load and making the
  system more prone to collisions. In this paper, we leverage machine learning
  techniques to design a system that outperforms the state-of-the-art schemes in
  preamble detection for the \gls{lte} \gls{rach} procedure. Most importantly,
  our scheme can also estimate the collision multiplicity, and thus gather
  information about how many devices chose the same preamble. This data can be
  used by the \gls{enb} to resolve collisions, increase the supported system
  load and reduce transmission latency. The presented approach is applicable to
  novel 3GPP standards that target massive IoT, e.g., LTE-M and NB-IoT.
\end{abstract}

\glsresetall

\section{Introduction}
\label{sec:introduction}

The \gls{rach} procedure in \gls{lte} serves as a synchronization
mechanism between the \gls{enb} and the \glspl{ue}, where the \glspl{ue} compete for
resources by randomly choosing a preamble and getting assigned data resources by
the \gls{enb}, provided that they do not collide with each other in the
preamble-based contention phase.

The current system was designed to handle about 128 attempts per
second~\cite{sesia2011lte}, however forecasts show that traffic could reach
upwards of 370 attempts per second in the near
future~\cite{madueno2014reengineering}, mainly due to the expected increase of the machine-type traffic in the cellular access. In case of
synchronized alarms, simultaneous accesses to the channel may yield up to a
tenfold increase with respect to normal traffic, almost guaranteeing that all
\glspl{ue} will collide when accessing the channel~\cite{cheng2012overload,
  laya2014is}. \gls{rach} overload represents a serious issue that can
significantly degrade the network performance, and in the last years a variety
of approaches to mitigate collisions were proposed. Studies generally focus either on improving the preamble detection or on how to efficiently use the random access-related resources. 

In this work, we propose a new machine learning based detection scheme that
requires no modification of the current protocol stack, and that can be fully
implemented at the \gls{enb}. This scheme outperforms the currently used
threshold based detection algorithms described in~\cite{sesia2011lte,
  lopez2012performance}, which aim at identifying whether a preamble has been
chosen by any \gls{ue} or not. Moreover, we have found machine learning
techniques to be particularly good at estimating the number of \glspl{ue}
picking the same preamble, providing information about the collision that is not
made available with the current techniques. Such information could be used to
tune successive stages of the contention resolution phase, thereby handling even
higher traffic loads.

The rest of the paper is organized as follows. Section~\ref{sec:lte_rach}
describes the LTE \gls{rach} procedure and the currently used algorithms for
preamble detection. Section~\ref{sec:ml} explains the dataset generation and the
machine learning approaches that were used. Section~\ref{sec:results} shows the
improvements brought by the new scheme with respect to current approaches, and
explores new features that can be enabled by collision multiplicity detection.
Finally, Section~\ref{sec:conclusion} concludes the paper.

\section{LTE RACH}
\label{sec:lte_rach}

The random access procedure in \gls{lte} is performed in the \gls{prach}, a
dedicated physical channel with an overall bandwidth of 1.08 MHz and duration
between 1 and 4 \gls{lte} subframes. We first describe how \gls{prach} preambles
are generated, then follow up the discussion with how they are used in the
multiple channel access scheme, and finally give an overview of the
state-of-the-art algorithms for preamble detection.

\subsection{Preamble generation} \label{sec:preamble}

LTE uses \gls{zc} sequences, complex-valued sequences that satisfy the Constant
Amplitude Zero Autocorrelation (CAZAC) property, as a basis to create RACH
preambles. A \gls{zc} sequence of odd length $N_{\textrm{ZC}}$ is defined as:
\begin{equation}
  z_r(n) = \exp\left\{-j 2 \pi r \frac{n(n+1)}{N_{\textrm{ZC}}}\right\},\quad n =0,1,\dots,N_{\textrm{ZC}}\!-\!1
\end{equation}
where $r\in \{1,\dots,N_{\textrm{ZC}}\!-\!1\}$ is the sequence root index; the
\gls{lte} standard uses $N_{\textrm{ZC}}=839$\footnote{Except for preamble
  format 4, which is not treated in this paper.}.
Given a root $r$, it is possible to generate multiple versions of the base
sequence $z_r$ through a circular shift, thereby obtaining orthogonal sequences
that exhibit a zero correlation with one another at the receiver. The cyclic
correlation of a \gls{zc} sequence with its root is a delta function with a
peak corresponding to the circular shift that was applied to the root sequence. It follows that circular correlation of a root against a
superposition of different sequences obtained from that root results in multiple
peaks that correspond to the individual shifts.

According to the cell size, the \gls{prach} preamble can have four different
formats, with duration from 1 to 4 subframes. In this work, we consider the
preamble format $0$,
which is typically used in cells with a radius up to 14 km, and consists in a
normal 1 ms random access burst with preamble sequences of duration $800$
$\mu$s.\footnote{We refer the reader to~\cite{sesia2011lte} for an exhaustive
 description of the different preamble formats and their structure.} The full
\gls{prach} preamble is obtained by prepending the so called \gls{cp} to the
\gls{zc} sequence chosen by the \gls{ue}. The \gls{cp} is a replica of the last
few symbols of the sequence, that helps counteracting the multi-path reflection
delay spread, and whose length is specified according to the chosen preamble
format. Additionally, the cyclic prefix makes it so that if the preamble signal
is delayed in time (like in the case of a \gls{ue} at the cell edge), there is
an additional shift in the correlation peak. While this feature enables the
\gls{enb} to estimate the channel delay experienced by a device, it also means
that different preambles arriving with different delays can yield a peak in the
same location of the correlation signal. In order to maintain the separation of
different preamble correlation peaks in the presence of delays, \gls{lte}
\glspl{ue} are allowed to only pick sequences whose shift is a multiple of a
base quantity $N_{\textrm{CS}}$, which depends on the cell radius (and hence on
the maximum delay) and on the propagation profile. $N_{\textrm{CS}}$ is in fact
chosen in such a way that each cyclic shift, when viewed within the time domain
of the signal, is greater than the combined maximum round trip propagation time
and multi-path delay-spread. This guarantees that the \gls{enb} can identify
different preambles by applying a correlator and a peak detector to the received
signal.

The cyclic cross-correlation between any two \gls{zc} sequences generated from
different roots is instead a constant value. Hence, in order to reduce noise at
the correlation, it is preferable to generate all preambles using as few root
indices as possible. Each root allows to obtain $\lfloor
N_{\textrm{ZC}}/N_{\textrm{CS}}\rfloor$ different preambles, and the \gls{lte}
standard contemplates the use of $N_{\textrm{prb}} = 64$ preambles in each cell.
By assigning orthogonal \gls{zc} sequences to adjacent \glspl{enb}, the
inter-cell interference is highly reduced. In this work, we only focus on the
case $N_{\textrm{CS}} = 13$, which is the smallest possible cyclic shift
dimension that allows to obtain exactly $N_{\textrm{prb}}$ preambles, so that a
single root is used to generate all the \gls{lte} preambles of the cell. This
configuration can be used in all cells covering a radius smaller than
0.79~km~\cite{sesia2011lte}, and is thus well suited to represent densely
deployed urban scenarios.



\subsection{Procedure}

\begin{figure}[t]
  \centering
  \includegraphics[width=.8\linewidth]{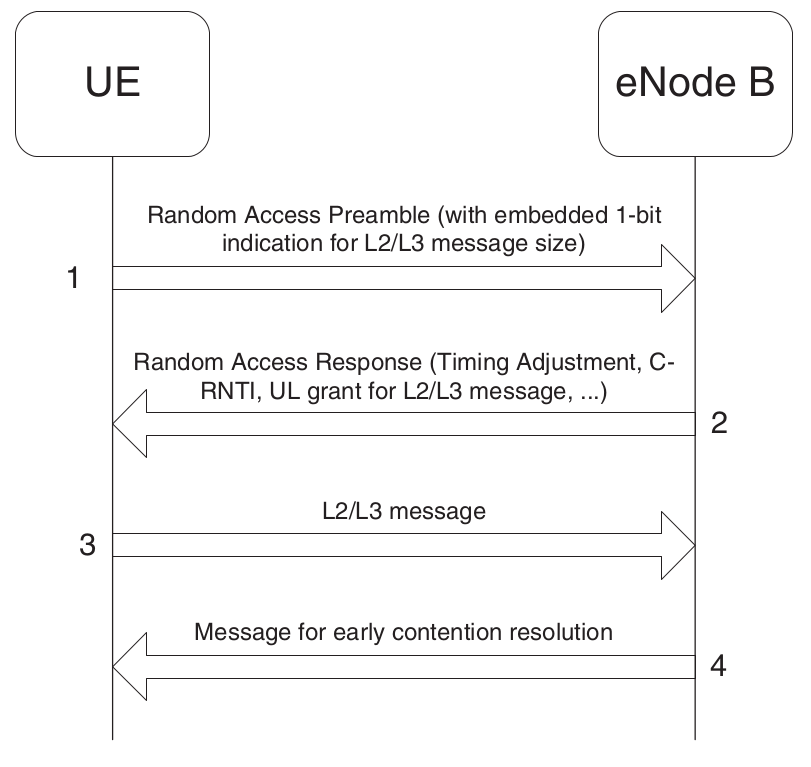}
  \caption{The LTE RACH procedure~\cite{sesia2011lte}.}
  \label{fig:rach_procedure}
\end{figure}

The \gls{rach} procedure consists of four phases (see
Fig.~\ref{fig:rach_procedure}):

\begin{enumerate}
\item \emph{Random Access Preamble}. The \glspl{ue} that intend to transmit data
  randomly choose one among $N_{\textrm{prb}}$ available preambles and send
  it to the \gls{enb}.

\item \emph{Random Access Response (RAR)}. The \gls{enb} processes the received
  signal, consisting in the superposition of all the transmitted preambles, and
  detects which preambles were chosen. For each detected
  preamble, it then sends a RAR message to assign the uplink resources to the
  corresponding \gls{ue}.

\item \emph{L2/L3 message}. The \glspl{ue} use the newly assigned data channel
  resources to communicate their connection request, and a unique identifier.

\item \emph{Contention Resolution Message}. The \gls{enb} responds to the
  \glspl{ue} using the identifiers they communicated in their L2/L3 message,
  granting the requested resources.
\end{enumerate}

\glspl{ue} can access the channel either in a \emph{contention-free} mode, where
the \gls{enb} forces the \glspl{ue} to use a particular signature for the
preamble generation, thus avoiding collisions, or in a \emph{contention-based}
mode. While contention-free access is reserved for handover and other special
delay-sensitive cases, contention-based access represents the default method for
\glspl{ue} to access the channel, and is thus taken as the focus of this paper.
In this procedure, which is illustrated in Figure~\ref{fig:rach_procedure}, if
multiple \glspl{ue} pick the same preamble, a collision happens, and may be
resolved at different stages of the procedure. If the colliding preambles are
received with high enough \gls{snr}, and are sufficiently spaced apart in
time\footnote{More specifically, the energy peaks corresponding to the preambles
  should be separated, in time, by at least the maximum delay spread of the
  cell.}, the \gls{enb} can detect both of them and recognize the collision, avoiding to send the RAR message to the involved \glspl{ue}.
  Otherwise, the \gls{enb} will not detect the collision but a single preamble, and all colliding \glspl{ue} will receive a
RAR message and try to access the same resource simultaneously,
colliding in the L2/L3 message phase. The \gls{enb} will therefore be unable
to decode the received message, and will send no contention resolution message;
the collided \glspl{ue} can then try again in a new \gls{prach} phase.
Collisions go undetected especially in urban environments, where smaller cells
are employed and the distance between different \glspl{ue} is too small to allow a differentiation of multiple copies of the
same preamble through the delay with which they arrive at the
\gls{enb}~\cite{laya2014is}.

\setlength{\abovecaptionskip}{5pt plus 2pt minus 0pt}

\subsection{State of the art in preamble detection} \label{sec:sota_detection}

The conventional approach to preamble detection in LTE RACH is provided
in~\cite{sesia2011lte}. This category of detectors (which we will refer to as \emph{threshold-based}) work by comparing the circular correlation of the received signal with its base sequence against a previously set threshold. A preamble is detected when the corresponding bin in the correlation signal contains values above the threshold. Such threshold is typically a function of the estimated noise level: this provides direct control over the false alarm probability, which can be made arbitrarily low (at a price in missed detection performance).

The threshold mechanism described previously is extended
in~\cite{lopez2012performance} to take into account quantization and
discretization steps that can improve computational performance, at the expense
of missed detection probability. Another approach to increase the detection performance for LTE frequency division duplex systems is presented in~\cite{li2011effective}, where noise is smoothed through an additional preprocessing phase prior to computing the correlation signal. This improves the performance in an ideal \gls{awgn} channel and with low \gls{snr}, but may be impractical for fading channels.

Multiple root sequences are considered in~\cite{kim2017enhanced}, which proposes a preamble detector able to identify non-orthogonal preambles and suppress the noise rise. The key idea is that the \gls{enb} can detect preambles in an almost interference-free environment by  eliminating the interfering signals from the original received signal. The power delay profile (PDP) allows to obtain the channel profile, used by the \gls{enb} to reconstruct the preamble signal.

The problem of performance degradation due to time dispersion of the channel is investigated in~\cite{yang2013enhanced}. Under the assumption of known PDP of Rayleigh fading that is independent across antennas and multiple paths, the paper derives an optimal statistic for preamble detecion for frequency selective channel, as an alternative to increasing the \gls{enb} target \gls{prach} received power to counter the time dispersion. 

\subsection{Interest in multiplicity detection}

Here we briefly comment the benefits of the multiplicity detection.
A straightforward application is to avoid sending a RAR message when multiple \gls{ue}s activated the same preamble, so that the involved \gls{ue}s will be implicitly informed about the collision. This enables to avoid subsequent collisions of L2/L3 messages and to shorten the \gls{rach} procedure.
Further, multiplicity detection may allow to infer the current load of the \gls{lte} \gls{rach}, as well as trends regarding its changes.
This could help, e.g., a proper dimensioning of the resources dedicated to the \gls{prach}, or adjust the operation of the \gls{rach} procedure. Examples are the dynamic allocation algorithm \cite{R2104662}, and the dynamic access class barring \cite{DSMW2013}, respectively. 

Another line of works where the multiplicity knowledge could be beneficial are the ones which assume reengineering of LTE \gls{rach} procedure, e.g., grouping LTE preambles in codewords that could convey information to the \gls{enb} \cite{PTSP2012,PSMP2016} or use of advanced collision resolution algorithms \gls{rach} \cite{MSP2014,MPSP2015}, which are shown to lower the latency and increase the reliability and throughput of LTE \gls{rach}.

We conclude by noting that the focus of the paper is on the multiplicity detection, while its coupling with advanced LTE \gls{rach} algorithms is left for further work.



\section{Machine learning approaches}
\label{sec:ml}

The main contribution of this work is the application of \gls{ml} to preamble
detection (i.e., determining whether a certain preamble was sent or not) and
preamble multiplicity detection (i.e., determining how many devices sent the
same preamble) in \gls{lte}. The necessary premise for every \gls{ml} algorithm
is to have data available, which can be used to train these systems and make
them ``learn''. Thus, in this section, we first give some details on the dataset
we used, and then discuss the employed \gls{ml} techniques, namely \gls{lr} and
\gls{nn}.


\subsection{Dataset generation} \label{sec:dataset}

Research about multiplicity collision in the \gls{rach} procedure is quite
modest, and, to the best of our knowledge, there is no publicly available
dataset that relates the signal received at the \gls{enb} with the information
of the preambles chosen by each station. Hence, we generated our own dataset,
using the MATLAB LTE
module.\footnote{\url{https://www.mathworks.com/products/lte-system.html}} The
LTE System Toolbox\texttrademark$\,$ in fact provides standard-compliant
functions for the design, simulation, and verification of both the LTE and
LTE-Advanced communications systems, with models up to Release 12 of the
standard (at the time this article was written).

\subsubsection{Data labeling}
To run the \gls{ml} algorithms, we need a map between (i) the signal received at
the \gls{enb} at the end of Phase 1) of the \gls{rach} procedure, and (ii) the
number of \glspl{ue} that selected each preamble.
These will represent the input and output data of the algorithms, respectively.
Accordingly, each simulation consists of simultaneous random access attempts
from a certain number of \glspl{ue}, each choosing one of the $N_{\textrm{prb}}$
available preambles. Using the MATLAB LTE module, it is possible to extrapolate
the corresponding \emph{correlation} signal at the \gls{enb}, i.e., the output
of the cyclic correlation of the signal received at the \gls{enb} with respect
to the root index, as described in Section~\ref{sec:preamble}.

\begin{figure}[t]
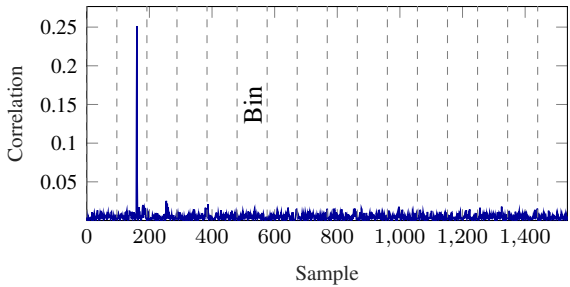

    \setlength\fwidth{0.9\columnwidth}
  \setlength\fheight{0.5\columnwidth}
  \include{correlation}
  \caption{Example of correlation signal with 16 bins.}
  \label{fig:correlation}
\end{figure}

As explained in Section~\ref{sec:preamble}, each preamble is uniquely mapped to
a correlation window of predefined size, according to the value of $N_{\textrm{CS}}$. We will denote such a window as \emph{bin}. Since we are
interested in how many devices chose each preamble, we decided to consider
each bin separately, rather than the correlation signal as a whole.
Figure~\ref{fig:correlation} shows an example output of the MATLAB LTE module.

\setlength{\abovecaptionskip}{-10pt plus 3pt minus 2pt}

\subsubsection{Applications}

We investigated two distinct applications:
\begin{itemize}
\item \emph{Preamble detection}, to compare the performance of the proposed
  \gls{ml} systems with respect to the state-of-the-art. For this application,
  the goal is to identify \emph{whether} a preamble was sent, and \emph{which}
  one. Three distinct datasets were generated to train and evaluate detection
  performance:
  \begin{itemize}
  \item A \emph{noise-only} set of correlations, used to evaluate false alarm
    probabilities according to the \gls{enb} testing
    specification~\cite{ts36.141};
  \item A set of correlation signals obtained when only a \emph{single UE} is
    transmitting, used to test missed detection;
  \item A dataset containing both \emph{noise-only and single UE} correlations,
    used to perform training of the \gls{ml} systems.
  \end{itemize}

  \item \emph{Multiplicity detection}, to derive the accuracy obtained when
  estimating the number of \glspl{ue} that chose the same preamble. For this purpose, we generated a dataset with bins containing from 0 to 5 preambles.
  This dataset was then split in two, for training and testing.
\end{itemize}

For both applications, we generated all datasets considering multiple scenarios,
which differ for:
\begin{itemize}
\item \emph{Noise level}. The signal received at the \gls{enb} is the
  superposition of all the transmitted preambles, corrupted by noise, expressed
  in terms of \gls{snr}. Based on performance from the \gls{ml} systems we
  tested, we decided to focus on an \gls{snr} range between -18 dB and -12
  dB. 

\item \emph{Channel model}. We considered an \gls{awgn} channel and an ETU70,
  i.e., an Extended Typical Urban channel with 70 Hz Doppler affected by
  Rayleigh fading.
\end{itemize}

For all scenarios, we ran over $10^5$ independent simulations, obtaining the
correlation signal in each bin and the number of \glspl{ue} that selected the
corresponding preamble. 

\subsubsection{Traffic intensity}
\label{sec:traffic}

While the preamble detection problem requires the transmission of at most one
preamble per \gls{rach} attempt, the number of competing devices is a critical parameter that influences the performance of the
multiplicity detection procedure. For our tests, we considered high traffic
scenarios corresponding, e.g., to alarm events that trigger transmission from a
large number of \glspl{ue}.

In~\cite{tr37.868}, 3GPP proposes a model for highly
correlated traffic arrivals, where the number of \glspl{ue} in each random
access opportunity inside the considered time frame follows a Beta distribution.
The standard~\cite{tr37.868} defines a possible massive access scenario with a
maximum of 30000 devices accessing the channel ``synchronously'', i.e., over a
time interval of about 10 s. Considering \gls{rach} opportunities to happen
every 20 ms, this corresponds to a maximum expected number of devices that
participate in the same \gls{rach} opportunity of about 120. We therefore generated our dataset with 120 \glspl{ue} that randomly choose
among the $N_{\textrm prb}$ available preambles. Due to the binomial
distribution nature of the preamble selection mechanism, 99\% of the bins
generated in this way have at most 5 devices choosing that preamble.

\subsection{Logistic regression}
\label{sec:logistic-regression}

\gls{lr} is a statistical method predicting the probability that a given input
data belongs or not to a certain class. %
The rationale behind \gls{lr} is that the input space can be separated into two
complementary regions, one for each class, by a linear boundary. It differs from
linear regression because the dependent, output variable is binary rather than
continuous, and this method is in fact intended to model classification
problems.
The probability that the output belongs to one or the other class is expressed
by means of a sigmoid $\sigma(\cdot)$, also called logistic function, from which
\gls{lr} takes its name.

In order to make a prediction, it is first necessary to \emph{train}
\gls{lr}, i.e., use known labeled data to determine the weights in
$\sigma(\cdot)$ that minimize a predefined cost function; such cost function
measures the distance between the desired output and the predicted one.

Given an \gls{snr} value, we trained different \gls{lr} models for each type of
channel considered (\gls{awgn} and Rayleigh), and for both the investigated
problems (preamble detection and collision multiplicity); the implementation was
done using the open source scikit-learn library in
Python\footnote{\url{https://scikit-learn.org}}. We used the dataset described
in Section~\ref{sec:dataset}, so that the logistic regressor is fed with the
correlation signal obtained at the receiver in a bin, while the output data is
the number of \glspl{ue} that picked the preamble corresponding to the
considered bin. For preamble detection, there are two possible classes, i.e., 0
or at least one \gls{ue}, while the number of possible classes in the collision
multiplicity problem is $N_{\max}+1$, where we set $N_{\max}=5$ as the maximum
number of colliding \glspl{ue} we are interested to estimate (see Section~\ref{sec:traffic}). In our case, choosing among
$N_{\max}+1$ alternatives can be modeled as a set of $N_{\max}$ independent
binary choices (a pivot alternative is compared to the remaining $N_{\max}$
ones). This makes the training complexity linear in
$N_{\max}$.

As typically done in classification problems, we gauged the performance of the
\gls{lr} predictors through the \emph{accuracy} metric, which represents the
percentage of correct predictions.


\subsection{Neural network}

Artificial \glspl{nn}~\cite{ding1996neural} are a class of mathematical models that are considered universal
approximators~\cite{hornik1991approximation}, as they are able to represent, up
to any accuracy, any non linear
function.
The basic units of \glspl{nn} are called \emph{neurons} and are organized in
multiple layers. Any \gls{nn} has an input layer fed with the data to
process, an output layer that represents the corresponding output determined
by the network, and possibly one or more hidden layers. Neurons represent
non-linear multi-input single-output functions; such functions are characterized
by some weights and biases, which represent the interconnections among the
neurons. Similarly to \gls{lr}, a \gls{nn} learns the relation between an
input data and its corresponding output through a training phase, during which
all the weights and biases are progressively tuned to produce the desired
output. \glspl{nn} are an extremely powerful tool because they can infer even
very complex functions, that analytical models or simpler approaches such as
\gls{lr} fail to describe.

We used a simple feedforward \gls{nn}, i.e., a fully-connected network where any
neuron in layer $i$ is connected to any neuron in the next layer $i+1$. For each
considered \gls{snr} value, channel type and problem to solve (preamble
detection or collision multiplicity), we trained a different \gls{nn} using the
Keras\footnote{\url{https://keras.io}} library.
Since our goal consists in determining the number of users that picked a certain
preamble, the input data is the correlation signal obtained at the \gls{enb} for
the corresponding bin. 
The output layer has $N_{\max}+1$ nodes, where $N_{\max}$ is the maximum
number of colliding \glspl{ue} we are interested in estimating ($N_{\max}=1$ for
the preamble detection, $N_{\max}=5$ for the collision
multiplicity).

In both scenarios, for all layers but the last one, the activation function
(i.e., the function that dictates how a neuron's weighted input is mapped into
its output) is the rectifier function, whereas for the output layer we decided
to use a softmax function, as usually done in classification problems. This
implies that the \gls{nn} outputs a separate probability for each of the
possible classes and chooses the most probable class. We finally evaluated the
efficiency of the neural network by means of accuracy, as we did for
\gls{lr}.

\section{Results}
\label{sec:results}

We gauged the performance of the \gls{lr} and \gls{nn} methods for preamble
detection and collision multiplicity estimation. In this section, we describe
the results we obtained and then discuss the new opportunities opened up by the
\gls{ml} tools, as well as their limitations.

\subsection{Preamble detection performance}

\begin{figure}[!t]
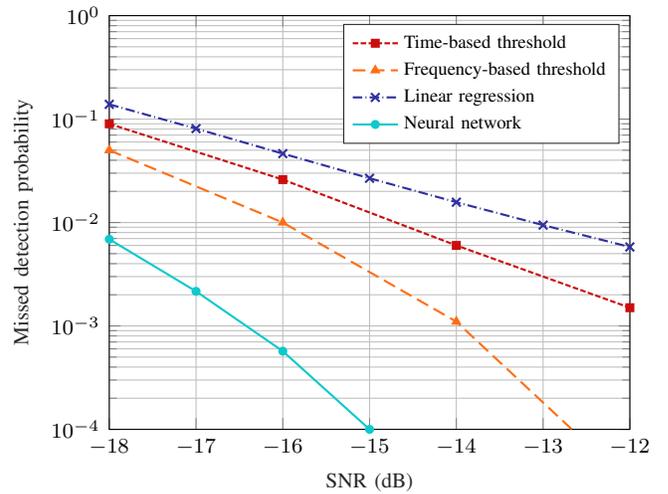

    \setlength\fwidth{0.96\columnwidth}
  \setlength\fheight{0.8\columnwidth}
  \centering
  \include{missed_detection_probability}
  \caption{\label{fig:missed_detection} Detection performance comparison in an
    AWGN channel.}
\end{figure}

\begin{figure}[t]
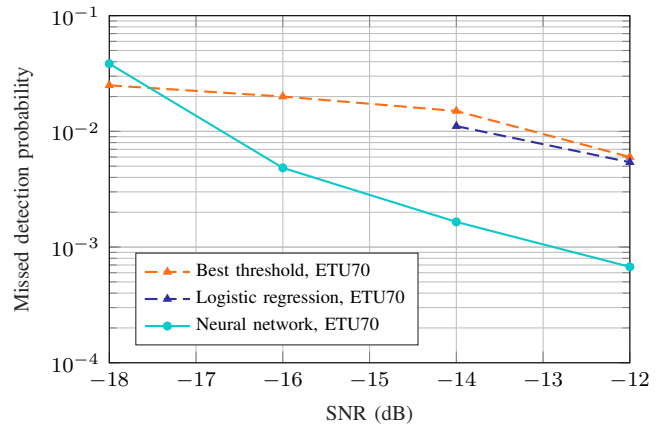

    \setlength\fwidth{0.96\columnwidth}
  \setlength\fheight{0.7\columnwidth}
  \centering
  \include{missed_detection_probability_awgn_rayleigh}
  \caption{\label{fig:missed_detection_awgn_rayleigh} \gls{nn} and best
    threshold system detection performance comparison in an ETU70 channel.}
\end{figure}

To guarantee a fair
comparison with the state-of-the-art threshold based detection, we conducted our
simulations according to the LTE specifications on \gls{enb} conformance
testing~\cite{ts36.141}, measuring the \emph{missed detection probability},
i.e., the probability that a transmitted preamble remains undetected at the
\gls{enb}. In this case, the test dataset (see Section~\ref{sec:dataset}) contains bins belonging to correlation signals in which exactly one device
was transmitting.  The dataset used for training of the \gls{ml} systems, instead, consists of bins coming from $10^5$ correlation signals
containing either 0 or 1 preambles.

The results for both the \gls{lr} and the \gls{nn} schemes in the case of an
\gls{awgn} channel are shown in Figure~\ref{fig:missed_detection}, together with
the performance of the algorithms proposed in~\cite{lopez2012performance} (see
Section~\ref{sec:sota_detection}). The scheme based on \gls{lr} yields worse
performance than those leveraging thresholds, because of its extreme simplicity;
the mediocre performance obtained with \gls{lr} suggests that the correlation
signal at the \gls{enb} and the transmission of defined preambles are data that
are not completely separable, but are rather related in a more complex way. This
is supported by the outstanding performance of the \gls{nn}, which provides a
gain in the range of 2 to 3 dB with respect to the best threshold-based detector
for all values of \gls{snr}, yielding a performance which conforms to the
standard requirements.
Rigorously, missed
detection probability should also include the cases of a wrong \gls{ue} delay
estimation. However, in the case of \gls{ml} based systems, which are not
capable of estimating this parameter, delay detection cannot be taken into
account.

Figure~\ref{fig:missed_detection_awgn_rayleigh} shows a comparison between the
detection performance of the \gls{nn} in the case of an ETU70 channel, and
compares it with the performance of the best threshold-based detection scheme.
Also in this case, the \gls{nn} achieves a significant improvement in detection
performance, upwards of 4 dB with respect to the threshold based schemes
described in~\cite{lopez2012performance}. \gls{lr} performance is only shown at
SNR values in which the false alarm probability requirement is satisfied.


A metric which is complementary to the missed detection probability is the
\emph{false alarm probability}, i.e., the probability that the \gls{enb} wrongly
detects a transmission in an unused bin. The minimum requirement for this metric
is set by the standard at $0.1\%$~\cite{ts36.141}. \gls{ml} approaches do not provide the same direct control over the false alarm probability that threshold-based schemes offer.
With the \gls{lr} scheme, the false alarm probability requirement is
respected for \gls{snr} larger than -16 dB. On the other hand, the false
alarm probability obtained with the \gls{nn} was consistently under the required $0.1\%$ threshold for all
the analyzed \gls{snr} values, for both AWGN and ETU70 channels.

\subsection{Multiplicity detection}

\begin{figure}[t]
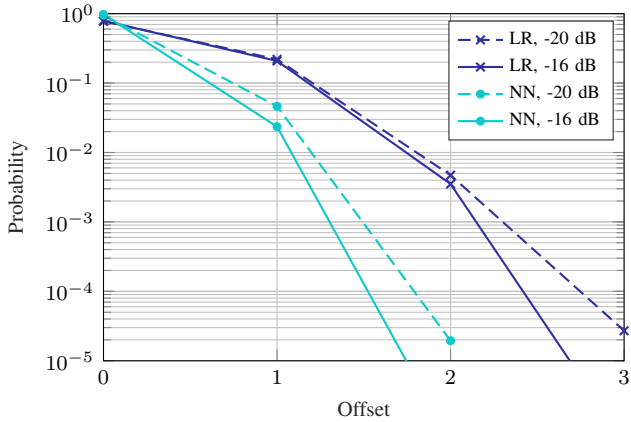

    \setlength\fwidth{0.96\columnwidth}
  \setlength\fheight{0.7\columnwidth}
  \centering
  \include{offset_probability}
  \caption{\label{fig:offset_probability} Probability of getting the
    multiplicity wrong by different offsets in an AWGN channel.}
\end{figure}

We assessed the reliability of the proposed schemes in the case of multiplicity
detection by measuring the frequency of errors in the estimation.
Figure~\ref{fig:offset_probability} shows the probabilities for both \gls{lr}
and \gls{nn} approaches to give a guess that is wrong by some amount, denoted as
offset. An offset of 0 represents the probability of guessing exactly the number
of preambles in the bin, an offset of 1 represents the probability that
the difference between the estimated and real number of transmitted preambles is
$\pm 1$, and so on.

\begin{figure}[t]
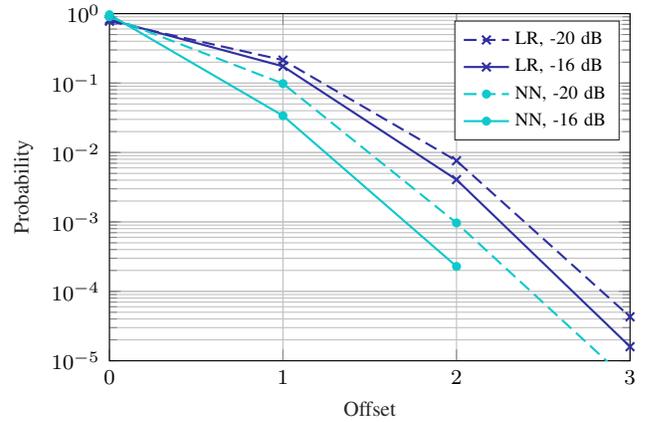

    \setlength\fwidth{0.96\columnwidth}
  \setlength\fheight{0.7\columnwidth}
  \centering
  \include{offset_rayleigh}
  \caption{\label{fig:offset_rayleigh} Probability of getting the
    multiplicity wrong by different offsets in an ETU70 channel.}
\end{figure}

As expected, lower \gls{snr} values give higher error probabilities. The
\gls{nn} clearly outperforms the \gls{lr} approach, thanks to its higher
complexity, which allows to infer even very complicated relations between its
input and output data. Although \gls{lr} could not compete with the \gls{nn}
in the preamble detection (see Figure~\ref{fig:missed_detection}), a linear
approach is still able to extract the information needed to identify the
number of transmitters in each bin from the signal received at the \gls{enb}
with a probability of around $0.8$, and gets the multiplicity wrong by at most 1
with a probability of $0.99$. The \gls{nn} scheme, instead, guarantees with a
probability of about $0.999$ to have an estimate that is either correct or only
off by one, even for very low \gls{snr} values, and yields exact guesses with a
probability of $0.9$.

Table~\ref{tab:confusion-matrix} contains the normalized confusion matrices for
the \gls{nn} and \gls{lr} systems, in the case of SNR $= -16$ dB: the rows
represent the actual multiplicity value, while the columns represent the value
estimated by the \gls{ml} scheme. Hence, the confusion matrix for an ideal
system would have ones along the diagonal, signifying that for each input the
labeling is correct with probability 1. It can be seen that the \gls{nn}
consistently outperforms the \gls{lr}, except for the case of no preambles being
sent, where \gls{lr} achieves a perfect score. Despite
this perfect performance in the correct detection of no preambles in the bin,
the \gls{lr} scheme is also heavily affected by a wrong detection rate in case
of 1 preamble being sent, thus motivating the bad overall detection performance
seen in Figure~\ref{fig:missed_detection}. Furthermore, the confusion matrices
also show that the \gls{lr} scheme tends to overestimate the number
of preambles in a bin, while the \gls{nn} yields a more symmetrical error.

\setlength{\abovecaptionskip}{5pt plus 10pt minus 0pt}
\begin{table}
  \centering
  \caption{Confusion matrices; SNR $=-16$ dB, AWGN channel.}\label{tab:confusion-matrix}
  \begin{tabular}{l|llllll}
    \toprule
    \multicolumn{7}{c}{Neural Network}                                                                      \\
    \midrule
      & 0              & 1              & 2              & 3              & 4              & 5              \\
    \midrule
    0 & \textbf{0.996} & 0.003          & 0              & 0              & 0              & 0              \\
    1 & 0              & \textbf{0.972} & 0.027          & 0              & 0              & 0              \\
    2 & 0              & 0.005          & \textbf{0.987} & 0.007          & 0              & 0              \\
    3 & 0              & 0              & 0.013          & \textbf{0.978} & 0.007          & 0              \\
    4 & 0              & 0              & 0              & 0.018          & \textbf{0.971} & 0.009          \\
    5 & 0              & 0              & 0              & 0              & 0.033          & \textbf{0.966} \\
    \toprule
    \multicolumn{7}{c}{Logistic regression}\\
    \midrule
      & 0              & 1              & 2              & 3              & 4              & 5              \\
    \midrule
    0 &\textbf{1}     & 0     &  0     &  0     & 0     &  0     \\
    1 &0.018 & \textbf{0.735} &  0.239 &  0.007 & 0     &  0     \\
    2 &0     & 0.048 &  \textbf{0.604} &  0.334 & 0.011 &  0     \\
    3 &0     & 0     &  0.062 &  \textbf{0.659} & 0.278 &  0     \\
    4 &0     & 0     &  0     &  0.086 & \textbf{0.813} &  0.099 \\
    5 &0     & 0     &  0     &  0     & 0.088 &  \textbf{0.912} \\
    \bottomrule
  \end{tabular}
\end{table}




\subsection{Machine learning limitations}

Although they have been shown to provide good results in preamble detection and collision
multiplicity estimation, \gls{ml} approaches entail some disadvantages, which we
describe next.

\subsubsection{Complexity}
even though the training phase of \gls{ml} based detection is performed offline,
such approaches are more complex than threshold based detection, and
inevitably require a larger computational power. In fact, while state-of-the-art
schemes only need to compare each correlation sample with a predefined
threshold, a detector based on \gls{lr} needs to perform a multiplication for
each sample, and then sum all the obtained results to get to a decision. To perform multiplicity
detection, such operations need to be performed for each
classifier as described in Section~\ref{sec:logistic-regression}. Neural
networks also need to perform a number of operations that is proportional to the
network complexity, and could be larger than the operations required by a
\gls{lr} scheme in the case of detection. When compared to a \gls{lr} scheme in
the case of multiplicity estimation, however, a \gls{nn} needs to perform
roughly the same number of operations as in the detection scheme, since it
already outputs a classification. 
In fairness to \gls{ml}, we remark that the complexity burden of the proposed
algorithms is on the \gls{enb}, whose hardware and software capabilities are
constantly improving and which typically have very little computational and
energy restrictions, such that the higher complexity demanded by \gls{ml}
schemes may not be a limiting factor.

\subsubsection{Dataset collection}
both in the case of \gls{lr} and \glspl{nn}, a dataset of sufficient size must
be available to perform effective training. In this work, the \gls{ml} based
approaches were trained on computer generated signals using a state-of-the-art
simulator in order to have a representation as close as possible to what the
\gls{enb} will actually see in a real deployment. This allowed us to exactly
know the number of \glspl{ue} choosing each preamble. For real deployments, this
dataset could be integrated with some samples taken by the \gls{enb}. In
particular, a bin associated to a certain preamble can be labeled according to
the outcome of the \gls{rach} procedure: label as 0 if no device sends a
\emph{msg3}, as 1 if exactly 1 device answers with a \emph{msg3}, and as 2 if a
collision happens during \emph{msg3}. Unfortunately, this does not allow for a
precise estimation of the number of devices sending a specific preamble.
Such an estimation may instead be possible if, once the \gls{enb} detects a
collision (i.e., if it labels the considered bin as 2), another ad-hoc contention
resolution phase were to take place iteratively among the colliding nodes, until
resolution of all collisions.


\section{Conclusions} \label{sec:conclusion}

With \gls{rach} overload becoming an increasingly serious issue in massive
access LTE scenarios, it is necessary to develop techniques to better manage the
contention of resources by multiple users. In this work, we have investigated
the use of machine learning techniques for preamble detection, and shown that
they can improve the performance of state-of-the-art threshold-based schemes. We
have also shown the capabilities of such techniques, and in particular of neural
networks, of inferring the number of users that pick a selected preamble, so as
to estimate the collision multiplicity.

The results described in this paper may serve as a first step to improve the
current \gls{rach} procedure, and are applicable to novel massive IoT 3GPP
standards, e.g., LTE-M and NB-IoT. In particular, the possibility to immediately
identify the number of devices that chose the same preamble allows more
efficient collision management than the current approach, which envisages a
repetition of the \gls{rach} procedure until all \glspl{ue} uniquely pick a
preamble. This would have a positive impact on the latency, the device power
consumption, and the supported system load. Moreover, collision multiplicity
detection may also allow to promptly identify a switch in the data reporting
regime, e.g., in case of an alarm that triggers synchronous transmissions from
multiple devices, and act accordingly.


\end{document}

%% file: correlation.tex
\definecolor{darkblue}{rgb}{0,0,0.6}%

\begin{tikzpicture}

\pgfplotsset{every tick label/.append style={font=\footnotesize}}

\begin{axis}[%
width=\fwidth,
height=\fheight,
xlabel style={font=\footnotesize\color{white!15!black}},
xlabel=Sample,
ylabel style={font=\footnotesize\color{white!15!black}},
ylabel=Correlation,
xmin=0,
xmax=1535,
yminorticks=true,
yminorgrids,
ymin=0,
ytick={0.05,0.1,0.15,0.2,0.25,0.3},
yticklabels={0.05,0.1,0.15,0.2,0.25,0.3},
]
\addplot [color=darkblue,solid,line width=0.6pt]
  table[row sep=crcr]{
    0 0.0030095\\
    1 0.00067685\\
    2 0.0060474\\
    3 0.003976\\
    4 0.001977\\
    5 0.0047426\\
    6 0.0033723\\
    7 0.0023453\\
    8 0.00057459\\
    9 0.002233\\
    10 0.0027944\\
    11 0.0033642\\
    12 0.0020248\\
    13 0.0023852\\
    14 0.0024839\\
    15 0.00061508\\
    16 0.0045931\\
    17 0.010888\\
    18 0.0052678\\
    19 0.00089705\\
    20 0.0039045\\
    21 0.0016294\\
    22 0.0010066\\
    23 0.0074432\\
    24 0.0099683\\
    25 0.0072768\\
    26 0.0048923\\
    27 0.0023311\\
    28 0.0010783\\
    29 0.003222\\
    30 0.0022762\\
    31 0.0046982\\
    32 0.0061696\\
    33 0.0022819\\
    34 0.013234\\
    35 0.0051244\\
    36 0.0055683\\
    37 0.0031061\\
    38 0.00049481\\
    39 0.0022897\\
    40 0.002906\\
    41 0.0044347\\
    42 0.0077334\\
    43 0.0040155\\
    44 0.0061014\\
    45 0.01254\\
    46 0.0049262\\
    47 0.0032195\\
    48 0.0027205\\
    49 0.0029579\\
    50 0.0068706\\
    51 0.0057902\\
    52 0.0082739\\
    53 0.0020027\\
    54 0.0037634\\
    55 0.0016558\\
    56 0.0015284\\
    57 0.0056877\\
    58 0.0038724\\
    59 0.004098\\
    60 0.0063141\\
    61 0.0065747\\
    62 0.0041241\\
    63 0.0043647\\
    64 0.0010594\\
    65 0.0096805\\
    66 0.0093384\\
    67 0.0077077\\
    68 0.0099969\\
    69 0.0036999\\
    70 0.0040813\\
    71 0.0045868\\
    72 0.0067751\\
    73 0.0054801\\
    74 0.0058786\\
    75 0.0049084\\
    76 0.0016319\\
    77 0.002577\\
    78 0.010568\\
    79 0.011212\\
    80 0.0067158\\
    81 0.0054006\\
    82 0.0027119\\
    83 0.004286\\
    84 0.003818\\
    85 0.001124\\
    86 0.0055544\\
    87 0.0028797\\
    88 0.00049051\\
    89 0.00096682\\
    90 0.0037013\\
    91 0.0064208\\
    92 0.0059491\\
    93 0.0051803\\
    94 0.0026834\\
    95 0.0035992\\
    96 0.0055968\\
    97 0.0045872\\
    98 0.0015942\\
    99 0.0027524\\
    100 0.0056492\\
    101 0.0022459\\
    102 0.0072861\\
    103 0.0057913\\
    104 0.0027627\\
    105 0.0013957\\
    106 0.00070506\\
    107 0.0058032\\
    108 0.0073371\\
    109 0.0084308\\
    110 0.0087256\\
    111 0.0043233\\
    112 0.0035528\\
    113 0.0029704\\
    114 0.0031107\\
    115 0.0057254\\
    116 0.0025564\\
    117 0.004042\\
    118 0.0047621\\
    119 0.0032182\\
    120 0.0029567\\
    121 0.00085995\\
    122 0.0049107\\
    123 0.010341\\
    124 0.010612\\
    125 0.00078722\\
    126 0.0041492\\
    127 0.010577\\
    128 0.012717\\
    129 0.0066396\\
    130 0.0052374\\
    131 0.0014475\\
    132 0.003005\\
    133 0.0014085\\
    134 0.0010308\\
    135 0.0059263\\
    136 0.012122\\
    137 0.011941\\
    138 0.0095961\\
    139 0.0084042\\
    140 0.0090244\\
    141 0.0063401\\
    142 0.0032263\\
    143 0.0040319\\
    144 0.0037833\\
    145 0.011245\\
    146 0.0059789\\
    147 0.0023804\\
    148 0.0048634\\
    149 0.0049961\\
    150 0.01085\\
    151 0.0030214\\
    152 0.0025611\\
    153 0.0017014\\
    154 0.0032128\\
    155 0.0086258\\
    156 0.0012587\\
    157 0.0082148\\
    158 0.004072\\
    159 0.14872\\
    160 0.25145\\
    161 0.051926\\
    162 0.0070458\\
    163 0.0044946\\
    164 0.0012478\\
    165 0.0056273\\
    166 0.017354\\
    167 0.001889\\
    168 0.011537\\
    169 0.011369\\
    170 0.0043221\\
    171 0.003492\\
    172 0.0071098\\
    173 0.0019637\\
    174 0.0035211\\
    175 0.0028122\\
    176 0.0054335\\
    177 0.010715\\
    178 0.016568\\
    179 0.020121\\
    180 0.0081043\\
    181 0.0037115\\
    182 0.011739\\
    183 0.012605\\
    184 0.016488\\
    185 0.014687\\
    186 0.0077294\\
    187 0.015896\\
    188 0.0092386\\
    189 0.0041457\\
    190 0.0048382\\
    191 0.0036114\\
    192 6.5787e-05\\
    193 0.0023426\\
    194 0.0029763\\
    195 0.011553\\
    196 0.010307\\
    197 0.0012019\\
    198 0.0015998\\
    199 0.0015817\\
    200 0.0014701\\
    201 0.0017925\\
    202 0.0010301\\
    203 0.0011051\\
    204 0.0034661\\
    205 0.0032305\\
    206 0.0020468\\
    207 0.0021075\\
    208 0.0015998\\
    209 0.0022432\\
    210 0.0074314\\
    211 0.0058817\\
    212 0.0050941\\
    213 0.0065886\\
    214 0.0064186\\
    215 0.0021054\\
    216 0.0033952\\
    217 0.0060724\\
    218 0.0046354\\
    219 0.0046114\\
    220 0.0012412\\
    221 0.0059323\\
    222 0.0098111\\
    223 0.0032729\\
    224 0.0011598\\
    225 0.0028991\\
    226 0.0092037\\
    227 0.0044515\\
    228 0.00036279\\
    229 0.0024854\\
    230 0.0052469\\
    231 0.0063306\\
    232 0.0063398\\
    233 0.0037877\\
    234 0.0018436\\
    235 0.0040916\\
    236 0.003573\\
    237 0.0010222\\
    238 0.0011612\\
    239 0.0027182\\
    240 0.00092539\\
    241 0.0026633\\
    242 0.0037643\\
    243 0.0023686\\
    244 0.0041682\\
    245 0.0036243\\
    246 0.0042604\\
    247 0.0080705\\
    248 0.0022933\\
    249 0.004323\\
    250 0.0016244\\
    251 0.00045863\\
    252 0.0016528\\
    253 0.017363\\
    254 0.025412\\
    255 0.010755\\
    256 0.0025008\\
    257 0.021728\\
    258 0.01134\\
    259 0.0058045\\
    260 0.016744\\
    261 0.0062975\\
    262 0.010242\\
    263 0.0021648\\
    264 0.0062592\\
    265 0.0044003\\
    266 0.003798\\
    267 0.00073323\\
    268 0.0029969\\
    269 0.0054002\\
    270 0.0050288\\
    271 0.0080112\\
    272 0.003678\\
    273 0.0028268\\
    274 0.0035354\\
    275 0.0069211\\
    276 0.0003897\\
    277 0.0070255\\
    278 0.0008878\\
    279 0.0026815\\
    280 0.0023511\\
    281 0.0017636\\
    282 0.003592\\
    283 0.0040718\\
    284 0.0032937\\
    285 0.0054408\\
    286 0.003131\\
    287 0.0026871\\
    288 0.0069369\\
    289 0.0042559\\
    290 0.0052748\\
    291 0.0081763\\
    292 0.0040788\\
    293 0.0014428\\
    294 0.0034286\\
    295 0.008827\\
    296 0.0082435\\
    297 0.0017613\\
    298 0.01032\\
    299 0.004621\\
    300 0.0022156\\
    301 0.002833\\
    302 0.0023077\\
    303 0.0056979\\
    304 0.0067344\\
    305 0.0031211\\
    306 0.0016968\\
    307 6.369e-05\\
    308 0.00015503\\
    309 0.0004217\\
    310 0.00076348\\
    311 0.0052896\\
    312 0.0028274\\
    313 0.0021589\\
    314 0.0062008\\
    315 0.0007676\\
    316 0.0026978\\
    317 0.0037425\\
    318 0.00060753\\
    319 0.0059447\\
    320 0.0058619\\
    321 0.0020947\\
    322 0.0049893\\
    323 0.003545\\
    324 0.0058602\\
    325 0.003882\\
    326 0.0076621\\
    327 0.0056769\\
    328 0.0032979\\
    329 0.0063067\\
    330 0.00039062\\
    331 0.0041836\\
    332 0.0058309\\
    333 0.0037653\\
    334 0.0053142\\
    335 0.0066521\\
    336 0.0025122\\
    337 0.0058987\\
    338 0.0015663\\
    339 0.0039553\\
    340 0.0048574\\
    341 0.0090839\\
    342 0.014408\\
    343 0.010233\\
    344 0.003397\\
    345 0.0019993\\
    346 0.0052396\\
    347 0.0009539\\
    348 0.0055533\\
    349 0.0063537\\
    350 0.0016014\\
    351 0.0020786\\
    352 0.0028434\\
    353 0.00070804\\
    354 0.0038074\\
    355 0.0057771\\
    356 0.0022429\\
    357 0.0066084\\
    358 0.0029041\\
    359 0.0043158\\
    360 0.0038616\\
    361 0.0061397\\
    362 0.0035877\\
    363 0.0040314\\
    364 0.004743\\
    365 0.0028409\\
    366 0.0032395\\
    367 0.0044502\\
    368 0.005783\\
    369 0.009689\\
    370 0.0051868\\
    371 0.00077784\\
    372 0.0025104\\
    373 0.0031211\\
    374 0.0028493\\
    375 0.0074107\\
    376 0.001422\\
    377 0.0051469\\
    378 0.0094574\\
    379 0.016563\\
    380 0.0047999\\
    381 0.0084014\\
    382 0.015246\\
    383 0.0077386\\
    384 0.010858\\
    385 0.0042956\\
    386 0.0050286\\
    387 0.021276\\
    388 0.0086413\\
    389 0.0034077\\
    390 0.0062346\\
    391 0.0043619\\
    392 0.00061046\\
    393 0.0070354\\
    394 0.0033425\\
    395 0.0029749\\
    396 0.0042063\\
    397 0.0023215\\
    398 0.004137\\
    399 0.0049084\\
    400 0.0042489\\
    401 0.0039179\\
    402 0.0012818\\
    403 0.0016136\\
    404 0.0018997\\
    405 0.0021632\\
    406 0.0016682\\
    407 0.0016013\\
    408 0.0012634\\
    409 0.0025083\\
    410 0.0043163\\
    411 0.0029875\\
    412 0.0038304\\
    413 0.0095782\\
    414 0.0096985\\
    415 0.0072809\\
    416 0.0062578\\
    417 0.0051869\\
    418 0.0031987\\
    419 0.0039357\\
    420 0.0099248\\
    421 0.0074656\\
    422 0.0013487\\
    423 0.0040913\\
    424 0.0044499\\
    425 0.001952\\
    426 0.0022167\\
    427 0.0044958\\
    428 0.0048717\\
    429 0.01338\\
    430 0.0077862\\
    431 0.0010507\\
    432 0.0075615\\
    433 0.0070338\\
    434 0.00044377\\
    435 0.0044032\\
    436 0.006762\\
    437 0.008012\\
    438 0.0060045\\
    439 0.0065731\\
    440 0.0061722\\
    441 0.0010369\\
    442 0.00059404\\
    443 0.0041133\\
    444 0.010628\\
    445 0.0053117\\
    446 0.0079624\\
    447 0.0045672\\
    448 0.0043232\\
    449 0.006144\\
    450 0.0062963\\
    451 0.0076428\\
    452 0.0041703\\
    453 0.00077593\\
    454 0.00042344\\
    455 0.0003542\\
    456 0.0017482\\
    457 0.0061621\\
    458 0.0059485\\
    459 0.00367\\
    460 0.010707\\
    461 0.0055022\\
    462 0.0027988\\
    463 0.012693\\
    464 0.0070798\\
    465 0.0016201\\
    466 0.0015472\\
    467 0.0032945\\
    468 0.0048834\\
    469 0.0022068\\
    470 0.0011728\\
    471 0.0073563\\
    472 0.0085423\\
    473 0.010203\\
    474 0.0065886\\
    475 0.0018388\\
    476 0.0049319\\
    477 0.0026868\\
    478 0.0025244\\
    479 0.0081542\\
    480 0.0052626\\
    481 0.0038842\\
    482 0.0010823\\
    483 0.0052793\\
    484 0.0040222\\
    485 0.00028948\\
    486 0.0013409\\
    487 0.0034177\\
    488 0.0022722\\
    489 0.00063355\\
    490 0.0043866\\
    491 0.007495\\
    492 0.0087334\\
    493 0.006946\\
    494 0.0067049\\
    495 0.00063357\\
    496 0.0051336\\
    497 0.0072583\\
    498 0.0031795\\
    499 0.011516\\
    500 0.0094557\\
    501 0.0043084\\
    502 0.0024439\\
    503 0.0060865\\
    504 0.0052955\\
    505 0.005142\\
    506 0.0064261\\
    507 0.0047837\\
    508 0.006584\\
    509 0.0088015\\
    510 0.0050241\\
    511 0.003921\\
    512 0.0037504\\
    513 0.0018191\\
    514 0.00064329\\
    515 0.0054761\\
    516 0.011923\\
    517 0.005972\\
    518 0.0049551\\
    519 0.01091\\
    520 0.0029138\\
    521 0.002914\\
    522 0.0036447\\
    523 0.0072272\\
    524 0.003949\\
    525 0.0035683\\
    526 0.0028282\\
    527 0.0034338\\
    528 0.0065512\\
    529 0.0056173\\
    530 0.0031832\\
    531 0.0015641\\
    532 0.0045437\\
    533 0.001859\\
    534 0.0099486\\
    535 0.012425\\
    536 0.0083547\\
    537 0.011686\\
    538 0.0069222\\
    539 0.0025183\\
    540 0.004063\\
    541 0.002779\\
    542 0.0032221\\
    543 0.0069309\\
    544 0.0037708\\
    545 0.0021303\\
    546 0.0031692\\
    547 0.0069297\\
    548 0.0096851\\
    549 0.0091779\\
    550 0.0078138\\
    551 0.0035275\\
    552 0.0032326\\
    553 0.0050245\\
    554 0.0062541\\
    555 0.0067359\\
    556 0.0023595\\
    557 0.0032386\\
    558 0.005889\\
    559 0.0042926\\
    560 0.0023405\\
    561 0.0088724\\
    562 0.0002969\\
    563 0.0055524\\
    564 0.0086958\\
    565 0.0064809\\
    566 0.0025383\\
    567 0.0025175\\
    568 0.0014439\\
    569 0.0017979\\
    570 0.0006343\\
    571 0.0022599\\
    572 0.0010154\\
    573 0.0033224\\
    574 0.0037107\\
    575 0.0044835\\
    576 0.0046533\\
    577 0.0010399\\
    578 0.0010933\\
    579 0.0013217\\
    580 0.0036116\\
    581 0.0022019\\
    582 0.0016395\\
    583 0.0020086\\
    584 0.0023688\\
    585 0.0057201\\
    586 0.0018364\\
    587 0.004636\\
    588 0.0024328\\
    589 0.00471\\
    590 0.0023706\\
    591 0.007519\\
    592 0.0067588\\
    593 0.0022844\\
    594 0.0021488\\
    595 0.004116\\
    596 0.0084174\\
    597 0.0074206\\
    598 0.003565\\
    599 0.0016577\\
    600 0.0039773\\
    601 0.00211\\
    602 0.0038073\\
    603 0.003009\\
    604 0.0074135\\
    605 0.0032648\\
    606 0.0036968\\
    607 0.0089559\\
    608 0.0059999\\
    609 0.0022301\\
    610 0.011488\\
    611 0.009447\\
    612 0.0034366\\
    613 0.0024531\\
    614 0.00092923\\
    615 0.0011582\\
    616 0.0035117\\
    617 0.0060305\\
    618 0.0069654\\
    619 0.0033975\\
    620 0.0086635\\
    621 0.010704\\
    622 0.0032001\\
    623 0.0014142\\
    624 0.0023971\\
    625 0.0040438\\
    626 0.00305\\
    627 0.0049695\\
    628 0.00020167\\
    629 0.0084297\\
    630 0.0041976\\
    631 0.0017414\\
    632 0.00041716\\
    633 0.0055922\\
    634 0.0056177\\
    635 0.0082733\\
    636 0.0061787\\
    637 0.0020576\\
    638 0.005051\\
    639 0.0011534\\
    640 0.0063663\\
    641 0.011326\\
    642 0.0062499\\
    643 0.016645\\
    644 0.005122\\
    645 0.0084819\\
    646 0.0055156\\
    647 0.0028978\\
    648 0.0030836\\
    649 0.00079148\\
    650 0.0056137\\
    651 0.0052339\\
    652 0.0017778\\
    653 0.00098279\\
    654 0.00025829\\
    655 0.0013011\\
    656 0.010521\\
    657 0.0064507\\
    658 0.0044261\\
    659 0.010319\\
    660 0.0031592\\
    661 0.00063154\\
    662 0.0008427\\
    663 0.00040016\\
    664 0.00056224\\
    665 0.0010431\\
    666 0.0012549\\
    667 0.0046601\\
    668 0.010784\\
    669 0.014193\\
    670 0.014499\\
    671 0.010424\\
    672 0.0096027\\
    673 0.0066898\\
    674 0.0039951\\
    675 0.0079989\\
    676 0.0067401\\
    677 0.0040264\\
    678 0.0044519\\
    679 0.00424\\
    680 0.0036625\\
    681 0.0026237\\
    682 0.0033155\\
    683 0.00025564\\
    684 0.0021252\\
    685 0.0019811\\
    686 0.0036179\\
    687 0.0048633\\
    688 0.0095893\\
    689 0.0068693\\
    690 0.0037858\\
    691 0.0048191\\
    692 0.0076903\\
    693 0.0027326\\
    694 0.0046627\\
    695 0.0037216\\
    696 0.0012477\\
    697 0.0043243\\
    698 0.010715\\
    699 0.0085485\\
    700 0.0017929\\
    701 0.0029649\\
    702 0.0012942\\
    703 0.0043407\\
    704 0.0058156\\
    705 0.0089362\\
    706 0.006393\\
    707 0.0034051\\
    708 0.0075156\\
    709 0.011665\\
    710 0.0052084\\
    711 0.0018742\\
    712 0.0088077\\
    713 0.0061539\\
    714 0.0098707\\
    715 0.0083437\\
    716 0.00084043\\
    717 0.0020486\\
    718 0.0099234\\
    719 0.0029812\\
    720 0.0040472\\
    721 0.0046053\\
    722 0.0081604\\
    723 0.010008\\
    724 0.0068534\\
    725 0.0060817\\
    726 0.0040954\\
    727 0.0013307\\
    728 0.0037032\\
    729 0.0022792\\
    730 0.0050381\\
    731 0.0031319\\
    732 0.0028504\\
    733 0.0036441\\
    734 0.0034638\\
    735 0.0068287\\
    736 0.0042448\\
    737 0.0018471\\
    738 0.0092657\\
    739 0.0043011\\
    740 0.0030662\\
    741 0.0038374\\
    742 0.0015286\\
    743 0.0033384\\
    744 0.0059778\\
    745 0.0010028\\
    746 0.0029741\\
    747 0.0032266\\
    748 0.0075127\\
    749 0.0044415\\
    750 0.003711\\
    751 0.00057721\\
    752 0.0025782\\
    753 0.0049905\\
    754 0.0076251\\
    755 0.0099775\\
    756 0.0089673\\
    757 0.00048284\\
    758 0.0050534\\
    759 0.0067537\\
    760 0.0032184\\
    761 0.0046697\\
    762 0.002971\\
    763 0.0055374\\
    764 0.010711\\
    765 0.0016412\\
    766 0.0029215\\
    767 0.010886\\
    768 0.015198\\
    769 0.0060951\\
    770 0.0044964\\
    771 0.0028407\\
    772 0.0011978\\
    773 0.0031367\\
    774 0.0035713\\
    775 0.0024701\\
    776 0.0027936\\
    777 0.005243\\
    778 0.0026884\\
    779 0.0059678\\
    780 0.0039632\\
    781 0.0067097\\
    782 0.0029252\\
    783 0.0043177\\
    784 0.005581\\
    785 0.0036915\\
    786 0.0033005\\
    787 0.006332\\
    788 0.01521\\
    789 0.0074355\\
    790 0.001615\\
    791 0.013901\\
    792 0.014326\\
    793 0.0018159\\
    794 0.003126\\
    795 0.0034281\\
    796 0.0047903\\
    797 0.010315\\
    798 0.0065013\\
    799 0.0027063\\
    800 0.0067288\\
    801 0.0056922\\
    802 0.0028793\\
    803 0.0032272\\
    804 0.0021627\\
    805 0.0052342\\
    806 0.0067893\\
    807 0.0068012\\
    808 0.0010866\\
    809 0.0060252\\
    810 0.0034382\\
    811 0.008028\\
    812 0.0044756\\
    813 0.0079935\\
    814 0.014567\\
    815 0.0057885\\
    816 0.0039035\\
    817 0.0055378\\
    818 0.0036535\\
    819 0.005196\\
    820 0.0006451\\
    821 0.0025016\\
    822 0.00066824\\
    823 0.0016476\\
    824 0.0011213\\
    825 0.0048067\\
    826 0.0053438\\
    827 0.010311\\
    828 0.0045192\\
    829 0.0015537\\
    830 0.0014252\\
    831 0.0041918\\
    832 0.00065125\\
    833 0.008109\\
    834 0.0023759\\
    835 0.0045319\\
    836 0.0054798\\
    837 0.0038934\\
    838 0.0030522\\
    839 0.0045016\\
    840 0.0028884\\
    841 0.0013446\\
    842 0.0025096\\
    843 0.0035345\\
    844 0.00063351\\
    845 0.0045568\\
    846 0.0052912\\
    847 0.0022745\\
    848 0.0038186\\
    849 0.0030441\\
    850 0.0048868\\
    851 0.0020528\\
    852 0.01058\\
    853 0.0094201\\
    854 0.01519\\
    855 0.0045588\\
    856 0.009757\\
    857 0.015241\\
    858 0.0045861\\
    859 0.011913\\
    860 0.006777\\
    861 0.0017557\\
    862 0.0022416\\
    863 0.0040944\\
    864 0.0031927\\
    865 0.0042288\\
    866 0.0009896\\
    867 0.001092\\
    868 0.00082557\\
    869 0.0018464\\
    870 0.0024898\\
    871 0.0010043\\
    872 0.010344\\
    873 0.0074601\\
    874 0.0029985\\
    875 0.0032147\\
    876 0.0026507\\
    877 0.015673\\
    878 0.014451\\
    879 0.0096312\\
    880 0.0062784\\
    881 0.0039386\\
    882 0.0074384\\
    883 0.0025496\\
    884 0.0040592\\
    885 0.0098177\\
    886 0.0034465\\
    887 0.00047884\\
    888 0.0027907\\
    889 0.0015132\\
    890 0.0034314\\
    891 0.0032441\\
    892 0.0042883\\
    893 0.0098357\\
    894 0.0029758\\
    895 0.00032565\\
    896 0.00094681\\
    897 0.0048228\\
    898 0.0044676\\
    899 0.0032638\\
    900 0.001465\\
    901 0.00062994\\
    902 0.0048805\\
    903 0.0085629\\
    904 0.010486\\
    905 0.0058409\\
    906 0.0028395\\
    907 0.0027473\\
    908 0.0023552\\
    909 0.0065818\\
    910 0.01011\\
    911 0.0098278\\
    912 0.0052558\\
    913 0.0029389\\
    914 0.0032671\\
    915 0.0035378\\
    916 0.0072863\\
    917 0.0070685\\
    918 0.0068674\\
    919 0.0024695\\
    920 0.002833\\
    921 0.0005102\\
    922 0.0023787\\
    923 0.0025078\\
    924 0.0071917\\
    925 0.0099717\\
    926 0.0078576\\
    927 0.0059443\\
    928 0.002038\\
    929 0.0024045\\
    930 0.0027388\\
    931 0.0077023\\
    932 0.0060517\\
    933 0.0023324\\
    934 0.013901\\
    935 0.0078628\\
    936 0.0082434\\
    937 0.0038746\\
    938 0.0016648\\
    939 0.0016149\\
    940 0.0033913\\
    941 0.0064409\\
    942 0.0088504\\
    943 0.0024554\\
    944 0.00065157\\
    945 0.0021828\\
    946 0.0097817\\
    947 0.0081707\\
    948 0.0026262\\
    949 0.0014079\\
    950 0.0024645\\
    951 0.0064289\\
    952 0.0032561\\
    953 0.0022469\\
    954 0.0036105\\
    955 0.0021722\\
    956 0.0035151\\
    957 0.0021878\\
    958 0.0059121\\
    959 0.0055884\\
    960 0.0044378\\
    961 0.0076724\\
    962 0.0040242\\
    963 0.0020721\\
    964 0.0077498\\
    965 0.012775\\
    966 0.0085066\\
    967 0.0058824\\
    968 0.0041121\\
    969 0.0023318\\
    970 0.005327\\
    971 0.0028396\\
    972 0.0030809\\
    973 0.0037497\\
    974 0.0012027\\
    975 0.0037384\\
    976 0.00069882\\
    977 0.0052699\\
    978 0.0072919\\
    979 0.004845\\
    980 0.0046005\\
    981 0.0036459\\
    982 0.0045316\\
    983 0.0057866\\
    984 0.0033362\\
    985 0.0023921\\
    986 0.0053079\\
    987 0.0055248\\
    988 0.0011813\\
    989 0.0044022\\
    990 0.0025117\\
    991 0.0053513\\
    992 0.008283\\
    993 0.0041349\\
    994 0.0073883\\
    995 0.0065715\\
    996 0.0073232\\
    997 0.017803\\
    998 0.0035264\\
    999 0.005932\\
    1000 0.0048398\\
    1001 0.0034041\\
    1002 0.0054158\\
    1003 0.0018572\\
    1004 0.00298\\
    1005 0.0057666\\
    1006 0.0015569\\
    1007 0.008192\\
    1008 0.013924\\
    1009 0.006774\\
    1010 0.0041927\\
    1011 0.0022413\\
    1012 0.0052702\\
    1013 0.0043249\\
    1014 0.0024158\\
    1015 0.0038428\\
    1016 0.0074295\\
    1017 0.006229\\
    1018 0.001005\\
    1019 0.0047479\\
    1020 0.0042939\\
    1021 0.0033667\\
    1022 0.0017986\\
    1023 0.0015776\\
    1024 0.0054111\\
    1025 0.001753\\
    1026 0.0049717\\
    1027 0.0052284\\
    1028 0.0011211\\
    1029 0.0083624\\
    1030 0.0061637\\
    1031 0.0042995\\
    1032 0.002167\\
    1033 0.005393\\
    1034 0.0035125\\
    1035 0.00016513\\
    1036 0.0017245\\
    1037 0.0019407\\
    1038 0.0025734\\
    1039 0.0019092\\
    1040 0.0092475\\
    1041 0.0065407\\
    1042 0.0097392\\
    1043 0.0045813\\
    1044 0.0018191\\
    1045 0.0042224\\
    1046 0.00086879\\
    1047 0.0046495\\
    1048 0.0088541\\
    1049 0.00037212\\
    1050 0.0032293\\
    1051 0.0012337\\
    1052 0.0064875\\
    1053 0.0093789\\
    1054 0.0077489\\
    1055 0.0032006\\
    1056 0.0021525\\
    1057 0.0024566\\
    1058 0.0065668\\
    1059 0.0026923\\
    1060 0.0026364\\
    1061 0.0022284\\
    1062 0.0016488\\
    1063 0.00057571\\
    1064 0.0065427\\
    1065 0.013867\\
    1066 0.010815\\
    1067 0.0047701\\
    1068 0.0039276\\
    1069 0.0093387\\
    1070 0.00064189\\
    1071 0.0042452\\
    1072 0.0031276\\
    1073 0.0052786\\
    1074 0.011227\\
    1075 0.0086716\\
    1076 0.0044001\\
    1077 0.0054626\\
    1078 0.00050564\\
    1079 0.0023345\\
    1080 0.003094\\
    1081 0.00030729\\
    1082 0.0014352\\
    1083 0.0019099\\
    1084 0.0065465\\
    1085 0.0017636\\
    1086 0.0031112\\
    1087 0.006087\\
    1088 0.011107\\
    1089 0.0058833\\
    1090 0.0012216\\
    1091 0.0018222\\
    1092 0.0021254\\
    1093 0.0032491\\
    1094 0.0032354\\
    1095 0.0030441\\
    1096 0.0046776\\
    1097 0.0054484\\
    1098 0.0088948\\
    1099 0.0019111\\
    1100 0.0074111\\
    1101 0.01252\\
    1102 0.0035802\\
    1103 0.0018046\\
    1104 0.00065627\\
    1105 0.00037381\\
    1106 0.00017762\\
    1107 0.00338\\
    1108 0.0066462\\
    1109 0.0047032\\
    1110 0.0018516\\
    1111 0.0003313\\
    1112 0.0031586\\
    1113 0.0039238\\
    1114 0.0023163\\
    1115 0.0041129\\
    1116 0.0030922\\
    1117 0.0062237\\
    1118 0.0017972\\
    1119 0.0029966\\
    1120 0.002836\\
    1121 0.0066726\\
    1122 0.005113\\
    1123 0.0014206\\
    1124 0.004769\\
    1125 0.0073075\\
    1126 0.0077525\\
    1127 0.011149\\
    1128 0.00053109\\
    1129 0.0075242\\
    1130 0.0068898\\
    1131 0.0064969\\
    1132 0.0077325\\
    1133 0.0033418\\
    1134 0.0027477\\
    1135 0.0083392\\
    1136 0.00037601\\
    1137 0.0054016\\
    1138 0.0050098\\
    1139 0.004245\\
    1140 0.00024387\\
    1141 0.0028422\\
    1142 0.0070846\\
    1143 0.0086935\\
    1144 0.0059231\\
    1145 0.0082954\\
    1146 0.0025814\\
    1147 0.00063834\\
    1148 0.0067767\\
    1149 0.010247\\
    1150 0.0018845\\
    1151 0.0017809\\
    1152 0.0050091\\
    1153 0.0040483\\
    1154 0.0039176\\
    1155 0.0049809\\
    1156 0.0028967\\
    1157 0.0033676\\
    1158 0.001525\\
    1159 0.0066142\\
    1160 0.0015945\\
    1161 0.0046207\\
    1162 0.0060471\\
    1163 0.002024\\
    1164 0.0020847\\
    1165 0.0099709\\
    1166 0.010371\\
    1167 0.0029147\\
    1168 0.0039721\\
    1169 0.0023509\\
    1170 0.0023035\\
    1171 0.0037313\\
    1172 0.0034673\\
    1173 0.0027006\\
    1174 0.0035316\\
    1175 0.0032335\\
    1176 0.00019482\\
    1177 0.0030111\\
    1178 0.00223\\
    1179 0.0019678\\
    1180 0.0024144\\
    1181 0.0055692\\
    1182 0.0027069\\
    1183 0.0017832\\
    1184 0.0033116\\
    1185 0.0093518\\
    1186 0.0044513\\
    1187 0.0040474\\
    1188 0.0020649\\
    1189 0.0077302\\
    1190 0.0061619\\
    1191 0.0022899\\
    1192 0.010132\\
    1193 0.011764\\
    1194 0.0047221\\
    1195 0.002028\\
    1196 0.002321\\
    1197 0.0025892\\
    1198 0.0032403\\
    1199 0.0024419\\
    1200 0.00055172\\
    1201 0.0004071\\
    1202 0.0021688\\
    1203 0.0035102\\
    1204 0.0045513\\
    1205 0.0035458\\
    1206 0.0033192\\
    1207 0.0068652\\
    1208 0.0048785\\
    1209 0.0044259\\
    1210 0.0064425\\
    1211 0.007703\\
    1212 0.013395\\
    1213 0.0081524\\
    1214 0.002062\\
    1215 0.0049166\\
    1216 0.0048733\\
    1217 0.0069811\\
    1218 0.0069591\\
    1219 0.0020236\\
    1220 0.006693\\
    1221 0.0044182\\
    1222 0.011938\\
    1223 0.01442\\
    1224 0.0023668\\
    1225 0.0018523\\
    1226 0.0017715\\
    1227 0.0038992\\
    1228 0.003506\\
    1229 0.0017802\\
    1230 0.00049504\\
    1231 0.0036157\\
    1232 0.0038659\\
    1233 0.0012758\\
    1234 0.00317\\
    1235 0.0047254\\
    1236 0.0025289\\
    1237 0.0017091\\
    1238 0.0041292\\
    1239 0.0014419\\
    1240 0.00036332\\
    1241 0.0027896\\
    1242 0.0097732\\
    1243 0.0064\\
    1244 0.0025984\\
    1245 0.0025768\\
    1246 0.0017286\\
    1247 0.0027119\\
    1248 0.0030297\\
    1249 0.0028958\\
    1250 0.0015944\\
    1251 0.0028984\\
    1252 0.0012164\\
    1253 0.002889\\
    1254 0.0080143\\
    1255 0.012537\\
    1256 0.016522\\
    1257 0.0066939\\
    1258 0.0019318\\
    1259 0.0070176\\
    1260 0.0063514\\
    1261 0.001781\\
    1262 0.0024556\\
    1263 0.0065298\\
    1264 0.0062983\\
    1265 0.0083844\\
    1266 0.0059135\\
    1267 0.0017684\\
    1268 0.0045875\\
    1269 0.0082629\\
    1270 0.0059864\\
    1271 0.010632\\
    1272 0.0059526\\
    1273 0.00074975\\
    1274 0.0046541\\
    1275 0.0041893\\
    1276 0.0073794\\
    1277 0.0023386\\
    1278 0.005753\\
    1279 0.0033576\\
    1280 0.0037941\\
    1281 0.0072724\\
    1282 0.0059151\\
    1283 0.0019918\\
    1284 0.0030504\\
    1285 0.0033681\\
    1286 0.0023538\\
    1287 0.0098485\\
    1288 0.0079962\\
    1289 0.003002\\
    1290 0.0081419\\
    1291 0.0035177\\
    1292 0.0031945\\
    1293 0.0052678\\
    1294 0.008617\\
    1295 0.006691\\
    1296 0.0029481\\
    1297 0.0033788\\
    1298 0.0075792\\
    1299 0.0090868\\
    1300 0.0059118\\
    1301 0.0048785\\
    1302 0.00596\\
    1303 0.0014546\\
    1304 0.0028545\\
    1305 0.00098161\\
    1306 0.0050448\\
    1307 0.0014009\\
    1308 0.0020181\\
    1309 0.0024857\\
    1310 0.00088014\\
    1311 0.0058442\\
    1312 0.010724\\
    1313 0.0040662\\
    1314 0.0028775\\
    1315 0.0028029\\
    1316 0.0047795\\
    1317 0.010303\\
    1318 0.0096172\\
    1319 0.0040528\\
    1320 0.0024074\\
    1321 0.0031296\\
    1322 0.010962\\
    1323 0.00577\\
    1324 0.0065815\\
    1325 0.017955\\
    1326 0.0035503\\
    1327 0.0034937\\
    1328 0.00277\\
    1329 0.0011212\\
    1330 0.0041063\\
    1331 0.0060884\\
    1332 0.0022003\\
    1333 0.0062335\\
    1334 0.005149\\
    1335 0.002031\\
    1336 0.0064279\\
    1337 0.0034801\\
    1338 0.0035609\\
    1339 0.00084146\\
    1340 0.0030287\\
    1341 0.0080374\\
    1342 0.0035319\\
    1343 0.0045899\\
    1344 0.0039099\\
    1345 0.0044921\\
    1346 0.0027002\\
    1347 0.0013759\\
    1348 0.0013349\\
    1349 0.012483\\
    1350 0.0076751\\
    1351 0.001659\\
    1352 0.0055239\\
    1353 0.0032522\\
    1354 0.004401\\
    1355 0.0054815\\
    1356 0.0037832\\
    1357 0.0064508\\
    1358 0.0050094\\
    1359 0.0028508\\
    1360 0.0010163\\
    1361 0.0020101\\
    1362 0.0016429\\
    1363 0.0068071\\
    1364 0.0066354\\
    1365 0.0021735\\
    1366 0.0024884\\
    1367 0.00064842\\
    1368 0.0023995\\
    1369 0.011392\\
    1370 0.0081475\\
    1371 0.0026532\\
    1372 0.0043343\\
    1373 0.0022043\\
    1374 0.0045497\\
    1375 0.0028847\\
    1376 0.0043034\\
    1377 0.0015003\\
    1378 0.01038\\
    1379 0.0075845\\
    1380 0.010081\\
    1381 0.012256\\
    1382 0.0025193\\
    1383 0.0034529\\
    1384 0.007021\\
    1385 0.0062472\\
    1386 0.0020126\\
    1387 0.0070727\\
    1388 0.0058368\\
    1389 0.0012705\\
    1390 0.0057355\\
    1391 0.0080166\\
    1392 0.0018381\\
    1393 0.0039855\\
    1394 0.0061904\\
    1395 0.0049385\\
    1396 0.0070149\\
    1397 0.002398\\
    1398 0.0035591\\
    1399 0.0095245\\
    1400 0.0064241\\
    1401 0.011457\\
    1402 0.0045482\\
    1403 0.0030761\\
    1404 0.0056803\\
    1405 0.0082111\\
    1406 0.0061799\\
    1407 0.00217\\
    1408 0.0016696\\
    1409 0.0019721\\
    1410 0.0044162\\
    1411 0.005023\\
    1412 0.0011959\\
    1413 0.0020951\\
    1414 0.0010294\\
    1415 0.00090061\\
    1416 0.0030564\\
    1417 0.0045107\\
    1418 0.0041859\\
    1419 0.0031221\\
    1420 0.0015097\\
    1421 0.0014225\\
    1422 0.0016359\\
    1423 0.0016054\\
    1424 0.001461\\
    1425 0.0036117\\
    1426 0.0064402\\
    1427 0.0039212\\
    1428 0.0031561\\
    1429 0.0032764\\
    1430 0.0080104\\
    1431 0.013833\\
    1432 0.0072311\\
    1433 0.0016415\\
    1434 0.0070369\\
    1435 0.0073827\\
    1436 0.0035494\\
    1437 0.0085211\\
    1438 0.008525\\
    1439 0.015393\\
    1440 0.0026685\\
    1441 0.00088983\\
    1442 0.0048308\\
    1443 0.006598\\
    1444 0.005093\\
    1445 0.00038535\\
    1446 0.0046916\\
    1447 0.0030161\\
    1448 0.0011861\\
    1449 0.0035703\\
    1450 0.004971\\
    1451 0.0029322\\
    1452 0.0047529\\
    1453 0.007413\\
    1454 0.0064523\\
    1455 0.0017254\\
    1456 0.0058302\\
    1457 0.0050923\\
    1458 0.008082\\
    1459 0.0027343\\
    1460 0.0014148\\
    1461 0.0021477\\
    1462 0.0091587\\
    1463 0.00825\\
    1464 0.0029309\\
    1465 0.0013599\\
    1466 0.0012062\\
    1467 0.0035093\\
    1468 0.00072932\\
    1469 0.0002631\\
    1470 0.0014184\\
    1471 0.0020661\\
    1472 0.0018809\\
    1473 0.0030724\\
    1474 0.0024943\\
    1475 0.012675\\
    1476 0.0054024\\
    1477 0.0051015\\
    1478 0.010747\\
    1479 0.0040144\\
    1480 0.00097794\\
    1481 0.0030488\\
    1482 0.0038228\\
    1483 0.0037221\\
    1484 0.0042576\\
    1485 0.0024878\\
    1486 0.0040745\\
    1487 0.0061149\\
    1488 0.0039692\\
    1489 0.0054043\\
    1490 0.0038459\\
    1491 0.0048888\\
    1492 0.0021114\\
    1493 0.0048747\\
    1494 0.0041437\\
    1495 0.0051781\\
    1496 0.0049224\\
    1497 0.0045004\\
    1498 0.0042051\\
    1499 0.0017108\\
    1500 0.004246\\
    1501 0.0085793\\
    1502 0.0065487\\
    1503 0.0088271\\
    1504 0.0029966\\
    1505 0.0021258\\
    1506 0.0041158\\
    1507 0.0015629\\
    1508 0.0021169\\
    1509 0.002866\\
    1510 0.0010585\\
    1511 0.0028625\\
    1512 0.0024854\\
    1513 0.0039615\\
    1514 0.0017989\\
    1515 0.0086182\\
    1516 0.0064532\\
    1517 0.0022897\\
    1518 0.0067278\\
    1519 0.0043382\\
    1520 0.0029751\\
    1521 0.0051729\\
    1522 0.0039646\\
    1523 0.0031469\\
    1524 0.0039722\\
    1525 0.0020475\\
    1526 0.004234\\
    1527 0.0065808\\
    1528 0.0053587\\
    1529 0.0077027\\
    1530 0.002287\\
    1531 0.00097703\\
    1532 0.0047826\\
    1533 0.001669\\
    1534 0.0028094\\
    1535 0.0014499\\
};

    \node[rotate=90] at (530,0.15) {Bin};

    \draw [dashed, gray, line width=0.4pt] (0,-1) -- (0,1);
    \draw [dashed, gray, line width=0.4pt] (96,-1) -- (96,1);
    \draw [dashed, gray, line width=0.4pt] (192,-1) -- (192,1);
    \draw [dashed, gray, line width=0.4pt] (288,-1) -- (288,1);
    \draw [dashed, gray, line width=0.4pt] (384,-1) -- (384,1);
    \draw [dashed, gray, line width=0.4pt] (480,-1) -- (480,1);
    \draw [dashed, gray, line width=0.4pt] (576,-1) -- (576,1);
    \draw [dashed, gray, line width=0.4pt] (672,-1) -- (672,1);
    \draw [dashed, gray, line width=0.4pt] (768,-1) -- (768,1);
    \draw [dashed, gray, line width=0.4pt] (864,-1) -- (864,1);
    \draw [dashed, gray, line width=0.4pt] (960,-1) -- (960,1);
    \draw [dashed, gray, line width=0.4pt] (1056,-1) -- (1056,1);
    \draw [dashed, gray, line width=0.4pt] (1152,-1) -- (1152,1);
    \draw [dashed, gray, line width=0.4pt] (1248,-1) -- (1248,1);
    \draw [dashed, gray, line width=0.4pt] (1344,-1) -- (1344,1);
    \draw [dashed, gray, line width=0.4pt] (1440,-1) -- (1440,1);

\end{axis}
\end{tikzpicture}%

%% file: missed_detection_probability.tex
\definecolor{darkviolet}{rgb}{0.29,0,0.52}
\definecolor{violet}{rgb}{0.84,0.23,0.84}%
\definecolor{darkgreen}{rgb}{0.08,0.5,0.08}%
\definecolor{lightgreen}{rgb}{0.3,0.9,0.2}%
\definecolor{darkorange}{rgb}{0.8,0.05,0.05}%
\definecolor{orange}{rgb}{1,0.43,0.08}%
\definecolor{darkerred}{rgb}{0.9,0,0}%
\definecolor{darkblue}{rgb}{0.19,0.19,0.63}%
\definecolor{lightblue}{rgb}{0.05,0.79,0.8}%

\begin{tikzpicture}

\pgfplotsset{every tick label/.append style={font=\footnotesize}}

\tikzstyle{dashed}= [dash pattern=on 7.5*0.8*0.8pt off 7.5*0.4*0.8pt]
\tikzstyle{dashed2}= [dash pattern=on 7.5*0.35*0.8pt off 7.5*0.15*0.8pt]
\tikzstyle{dashdotted} = [dash pattern=on 7.5*0.6*0.8pt off 7.5*0.2*0.8pt on \the\pgflinewidth off 7.5*0.2*0.8pt]

\begin{axis}[%
width=\fwidth,
height=\fheight,
xlabel style={font=\footnotesize\color{white!15!black}},
xlabel=SNR (dB),
ylabel style={font=\footnotesize\color{white!15!black}},
ylabel=Missed detection probability,
xmin=-18,
xmax=-12,
xmajorgrids,
yminorticks=true,
yminorgrids,
ymode=log,
ymin=10^-4,
ymax=1,
ymajorgrids,
legend entries={\scriptsize{Time-based threshold},
	\scriptsize{Frequency-based threshold},
  \scriptsize{Linear regression},
  \scriptsize{Neural network}
	},
legend style={legend cell align=left,align=left,draw=white!15!black}
]
\addplot [color=darkorange,dashed2,mark=square*, mark size=1.3,mark options=solid,line width=0.8pt]
  table[row sep=crcr]{
-18 9e-2\\
-16 2.6e-2\\
-14 6e-3\\
-12 1.5e-3\\
-10 3.5e-4\\
-8 3e-5\\
};
\addplot [color=orange,dashed,mark=triangle*, mark size=1.5,mark options=solid,line width=0.8pt]
  table[row sep=crcr]{
-18 5e-2\\
-16 1e-2\\
-14 1.1e-3\\
-12 3e-5\\
};
\addplot [color=darkblue,dashdotted,mark=x,mark options=solid,mark size=2.3,line width=0.8pt]
  table[row sep=crcr]{
 -18 0.1389\\
 -17 0.08085\\
 -16 0.0463375\\
 -15 0.026775\\
 -14 0.0157125\\
 -13 0.0094375\\
 -12 0.0058\\
 };
\addplot [color=lightblue,solid,mark=*,mark size=1.3,line width=0.8pt]
  table[row sep=crcr]{
 -18 0.00689\\
 -17 0.00216\\
 -16 0.00057\\
 -15 0.0001\\
 };
\end{axis}
\end{tikzpicture}%

%% file: missed_detection_probability_awgn_rayleigh.tex
\definecolor{darkviolet}{rgb}{0.29,0,0.52}
\definecolor{violet}{rgb}{0.84,0.23,0.84}%
\definecolor{darkgreen}{rgb}{0.08,0.5,0.08}%
\definecolor{lightgreen}{rgb}{0.3,0.9,0.2}%
\definecolor{darkorange}{rgb}{0.8,0.05,0.05}%
\definecolor{orange}{rgb}{1,0.43,0.08}%
\definecolor{darkerred}{rgb}{0.9,0,0}%
\definecolor{darkblue}{rgb}{0.19,0.19,0.63}%
\definecolor{lightblue}{rgb}{0.05,0.79,0.8}%

\begin{tikzpicture}

\pgfplotsset{every tick label/.append style={font=\footnotesize}}
\tikzstyle{dashed}= [dash pattern=on 7.5*0.8*0.8pt off 7.5*0.4*0.8pt]

\begin{axis}[%
width=\fwidth,
height=\fheight,
xlabel style={font=\footnotesize\color{white!15!black}},
xlabel=SNR (dB),
ylabel style={font=\footnotesize\color{white!15!black}},
ylabel=Missed detection probability,
xmin=-18,
xmax=-12,
xmajorgrids,
yminorticks=true,
yminorgrids,
ymode=log,
ymin=10^-4,
ymax=10^-1,
ymajorgrids,
legend entries={
  \scriptsize{Best threshold, ETU70},
  \scriptsize{Logistic regression, ETU70},
  \scriptsize{Neural network, ETU70}
	},
legend style={at={(0.05,0.05)}, anchor=south west, legend cell align=left, align=left, draw=white!15!black},
]
\addplot [color=orange,dashed,mark=triangle*, mark size=1.3,mark options=solid,line width=0.8pt]
  table[row sep=crcr]{
-18 2.5e-2\\
-16 2e-2\\
-14 1.5e-2\\
-12 6e-3\\
};

\addplot [color=darkblue,dashed,mark=triangle*, mark size=1.3,mark options=solid,line width=0.8pt]
  table[row sep=crcr]{
-14 0.0111\\
-12 0.0054125\\
};
\addplot [color=lightblue,solid,mark=*,mark size=1.3,line width=0.8pt]
  table[row sep=crcr]{
-18 0.0384875\\
-16 0.0048375\\
-14 0.00165\\
-12 0.000675\\
};
\end{axis}
\end{tikzpicture}%

%% file: offset_probability.tex
\definecolor{darkviolet}{rgb}{0.29,0,0.52}
\definecolor{violet}{rgb}{0.84,0.23,0.84}%
\definecolor{darkgreen}{rgb}{0.08,0.5,0.08}%
\definecolor{lightgreen}{rgb}{0.3,0.9,0.2}%
\definecolor{darkorange}{rgb}{0.8,0.05,0.05}%
\definecolor{orange}{rgb}{1,0.43,0.08}%
\definecolor{darkerred}{rgb}{0.9,0,0}%
\definecolor{darkblue}{rgb}{0.19,0.19,0.63}%
\definecolor{lightblue}{rgb}{0.05,0.79,0.8}%

\begin{tikzpicture}
\pgfplotsset{every tick label/.append style={font=\footnotesize}}
\tikzstyle{dashed}= [dash pattern=on 7.5*0.8*0.8pt off 7.5*0.4*0.8pt]

\begin{axis}[%
width=\fwidth,
height=\fheight,
xlabel style={font=\footnotesize\color{white!15!black}},
xlabel=Offset,
ylabel style={font=\footnotesize\color{white!15!black}},
ylabel=Probability,
xmin=0,
xmax=3,
xmajorgrids,
xtick={0,1,2,3},
yminorticks=true,
yminorgrids,
ymode=log,
ymin=10^-5,
ymax=1,
ymajorgrids,
legend entries={
  \scriptsize{LR, -20 dB},
  \scriptsize{LR, -16 dB},
  \scriptsize{NN, -20 dB},
  \scriptsize{NN, -16 dB},
	},
  legend style={legend cell align=left,align=left,draw=white!15!black},
  mark options={solid}
]
\addplot [color=darkblue,dashed,mark=x,mark options=solid,mark size=2.3,line width=0.8pt]
  table[row sep=crcr]{
    0 7.75625035e-01\\
    1 2.19635668e-01\\
    2 4.71188017e-03\\
    3 2.70697984e-05\\
    4 3.47048698e-07\\
};
\addplot [color=darkblue,solid,mark=x,mark options=solid,mark size=2.3,line width=0.8pt]
  table[row sep=crcr]{
    0 7.88048337e-01\\
    1 2.08432242e-01\\
    2 3.51872675e-03\\
    3 6.94097396e-07\\
};
\addplot [color=lightblue,dashed,mark=*,mark size=1.3,mark options=solid,line width=0.8pt]
  table[row sep=crcr]{
    0 9.53699306e-01\\
    1 4.62812500e-02\\
    2 1.94444444e-05\\
};
\addplot [color=lightblue,solid,mark=*,mark size=1.3,mark options=solid,line width=0.8pt]
  table[row sep=crcr]{
    0 9.76462153e-01\\
    1 2.35371528e-02\\
    2 6.94444444e-07\\
};
\end{axis}
\end{tikzpicture}%

%% file: offset_rayleigh.tex
\definecolor{darkviolet}{rgb}{0.29,0,0.52}
\definecolor{violet}{rgb}{0.84,0.23,0.84}%
\definecolor{darkgreen}{rgb}{0.08,0.5,0.08}%
\definecolor{lightgreen}{rgb}{0.3,0.9,0.2}%
\definecolor{darkorange}{rgb}{0.8,0.05,0.05}%
\definecolor{orange}{rgb}{1,0.43,0.08}%
\definecolor{darkerred}{rgb}{0.9,0,0}%
\definecolor{darkblue}{rgb}{0.19,0.19,0.63}%
\definecolor{lightblue}{rgb}{0.05,0.79,0.8}%

\begin{tikzpicture}
\pgfplotsset{every tick label/.append style={font=\footnotesize}}
\tikzstyle{dashed}= [dash pattern=on 7.5*0.8*0.8pt off 7.5*0.4*0.8pt]

\begin{axis}[%
width=\fwidth,
height=\fheight,
xlabel style={font=\footnotesize\color{white!15!black}},
xlabel=Offset,
ylabel=Probability,
ylabel style={font=\footnotesize\color{white!15!black}},
xmin=0,
xmax=3,
xmajorgrids,
xtick={0,1,2,3},
yminorticks=true,
yminorgrids,
ymode=log,
ymin=10^-5,
ymax=1,
ymajorgrids,
legend entries={
  \scriptsize{LR, -20 dB},
  \scriptsize{LR, -16 dB},
  \scriptsize{NN, -20 dB},
  \scriptsize{NN, -16 dB},
	},
  legend style={legend cell align=left,align=left,draw=white!15!black},
  mark options={solid}
]
\addplot [color=darkblue,dashed,mark=x,mark options=solid,mark size=2.3,line width=0.8pt]
  table[row sep=crcr]{
    0 7.77978272e-01\\
    1 2.14346764e-01\\
    2 7.60628261e-03\\
    3 4.30125430e-05\\
    4 2.56687757e-05\\
};
\addplot [color=darkblue,solid,mark=x,mark options=solid,mark size=2.3,line width=0.8pt]
  table[row sep=crcr]{
    0 8.21640859e-01\\
    1 1.74276418e-01\\
    2 4.06607282e-03\\
    3 1.59562660e-05\\
    4 6.93750694e-07\\
};
\addplot [color=lightblue,dashed,mark=*,mark size=1.3,mark options=solid,line width=0.8pt]
  table[row sep=crcr]{
    0 9.00778472e-01\\
    1 9.82500000e-02\\
    2 9.65972222e-04\\
    3 4.86111111e-06\\
    4 6.94444444e-07\\
};
\addplot [color=lightblue,solid,mark=*,mark size=1.3,mark options=solid,line width=0.8pt]
  table[row sep=crcr]{
    0 9.65905556e-01\\
    1 3.38659722e-02\\
    2 2.28472222e-04\\
};

\end{axis}
\end{tikzpicture}%

%% file: report.bbl
\begin{thebibliography}{10}
\providecommand{\url}[1]{#1}
\csname url@samestyle\endcsname
\providecommand{\newblock}{\relax}
\providecommand{\bibinfo}[2]{#2}
\providecommand{\BIBentrySTDinterwordspacing}{\spaceskip=0pt\relax}
\providecommand{\BIBentryALTinterwordstretchfactor}{4}
\providecommand{\BIBentryALTinterwordspacing}{\spaceskip=\fontdimen2\font plus
\BIBentryALTinterwordstretchfactor\fontdimen3\font minus
  \fontdimen4\font\relax}
\providecommand{\BIBforeignlanguage}[2]{{%
\expandafter\ifx\csname l@#1\endcsname\relax
\typeout{** WARNING: IEEEtran.bst: No hyphenation pattern has been}%
\typeout{** loaded for the language `#1'. Using the pattern for}%
\typeout{** the default language instead.}%
\else
\language=\csname l@#1\endcsname
\fi
#2}}
\providecommand{\BIBdecl}{\relax}
\BIBdecl

\bibitem{sesia2011lte}
S.~Sesia, M.~Baker, and I.~Toufik, \emph{{LTE - The UMTS Long Term Evolution:
  from theory to practice}}.\hskip 1em plus 0.5em minus 0.4em\relax John Wiley
  \& Sons, 2011.

\bibitem{madueno2014reengineering}
G.~C. Madueno, {\v{C}}.~Stefanovi{\'c}, and P.~Popovski, ``Reengineering
  {GSM/GPRS} towards a dedicated network for massive smart metering,'' in
  \emph{IEEE International Conference On Smart Grid Communications
  (SmartGridComm)}.\hskip 1em plus 0.5em minus 0.4em\relax IEEE, 2014, pp.
  338--343.

\bibitem{cheng2012overload}
M.~Y. Cheng, G.~Y. Lin, H.~Y. Wei, and A.~C.~C. Hsu, ``Overload control for
  machine-type-communications in {LTE-Advanced} system,'' \emph{{IEEE}
  Communications Magazine}, vol.~50, no.~6, pp. 38--45, June 2012.

\bibitem{laya2014is}
A.~Laya, L.~Alonso, and J.~Alonso-Zarate, ``{Is the Random Access Channel of
  LTE and LTE-A Suitable for M2M Communications? A Survey of Alternatives},''
  \emph{IEEE Communications Surveys Tutorials}, vol.~16, no.~1, pp. 4--16,
  First 2014.

\bibitem{lopez2012performance}
F.~J. L{\'o}pez-Mart{\'\i}nez, E.~del Castillo-S{\'a}nchez, E.~Martos-Naya, and
  J.~T. Entrambasaguas, ``{Performance evaluation of preamble detectors for
  3GPP-LTE physical random access channel},'' \emph{Digital Signal Processing},
  vol.~22, no.~3, pp. 526--534, 2012.

\bibitem{li2011effective}
P.~Li and B.~Wu, ``An effective approach to detect random access preamble in
  {LTE} systems in low {SNR},'' \emph{Procedia Engineering}, vol.~15, pp.
  2339--2343, 2011.

\bibitem{kim2017enhanced}
T.~Kim, I.~Bang, and D.~K. Sung, ``An enhanced {PRACH} preamble detector for
  cellular {IoT} communications,'' \emph{IEEE Communications Letters}, vol.~21,
  no.~12, pp. 2678--2681, 2017.

\bibitem{yang2013enhanced}
X.~Yang and A.~O. Fapojuwo, ``Enhanced preamble detection for {PRACH} in
  {LTE},'' in \emph{IEEE Wireless Communications and Networking Conference
  (WCNC)}.\hskip 1em plus 0.5em minus 0.4em\relax IEEE, 2013, pp. 3306--3311.

\bibitem{R2104662}
3GPP, ``{MTC simulation results with specific solutions},'' {3rd Generation
  Partnership Project (3GPP)}, TR {R2-104662}, Aug. 2010.

\bibitem{DSMW2013}
S.~Duan, V.~Shah-Mansouri, and V.~W.~S. Wong, ``{Dynamic access class barring
  for M2M communications in LTE networks},'' in \emph{IEEE Global
  Communications Conference (GLOBECOM)}, Dec 2013, pp. 4747--4752.

\bibitem{PTSP2012}
N.~K. Pratas, H.~Thomsen, C.~Stefanovic, and P.~Popovski, ``{Code-expanded
  random access for machine-type communications},'' in \emph{IEEE Globecom
  Workshops}, Dec 2012, pp. 1681--1686.

\bibitem{PSMP2016}
N.~K. Pratas, C.~Stefanovic, G.~C. Madueno, and P.~Popovski, ``{Random Access
  for Machine-Type Communication Based on Bloom Filtering},'' in \emph{IEEE
  Global Communications Conference}, Dec 2016, pp. 1--7.

\bibitem{MSP2014}
G.~C. Madueno, C.~Stefanovic, and P.~Popovski, ``{Efficient LTE access with
  collision resolution for massive M2M communications},'' in \emph{IEEE
  Globecom Workshops (GC Wkshps)}, Dec 2014, pp. 1433--1438.

\bibitem{MPSP2015}
G.~C. Madueno, N.~K. Pratas, C.~Stefanovic, and P.~Popovski, ``{Massive M2M
  access with reliability guarantees in LTE systems},'' in \emph{IEEE
  International Conference on Communications}, 2015, pp. 2997--3002.

\bibitem{ts36.141}
{3GPP}, ``{LTE; Evolved Universal Terrestrial Radio Access (E-UTRA); Base
  Station (BS)} conformance testing,'' TS 36.141 V.11.3.0.

\bibitem{tr37.868}
------, ``Study on {RAN} improvements for machine-type communications,'' TR
  37.868 V.11.0.0.

\bibitem{ding1996neural}
N.~K. Bose and P.~Liang, ``Neural network fundamentals with graphs, algorithms
  and applications,'' 1996.

\bibitem{hornik1991approximation}
K.~Hornik, ``Approximation capabilities of multilayer feedforward networks,''
  \emph{Neural networks}, vol.~4, no.~2, pp. 251--257, 1991.

\end{thebibliography}
